\documentclass[onecolumn,11.5pt]{article}
\usepackage[numbers,sort&compress]{natbib}
\usepackage[textwidth=6.5in,textheight=8.2in,headsep=80pt,footskip=65pt]{geometry}%

\usepackage{amsmath}
\usepackage{amssymb}

\usepackage{rotating}%rotate figures
\usepackage{url}
\usepackage{graphicx}% Include figure files
\usepackage{dcolumn}% Align table columns on decimal point
\usepackage{bm}% bold math
\usepackage[usenames,dvipsnames]{xcolor} % extra colors
\usepackage{stmaryrd} %bracket
\usepackage{bbm}

\usepackage{xcolor} % extra colors
\newcommand{\be}{\begin{equation}}
\newcommand{\ee}{\end{equation}}
\newcommand{\bea}{\begin{eqnarray}}
\newcommand{\eea}{\end{eqnarray}}
\newcommand{\beas}{\begin{eqnarray*}}
\newcommand{\eeas}{\end{eqnarray*}}

\newcommand{\ER}{Erd\H{o}s-R\'{e}nyi }

\begin{document}

\title{Network models of financial systemic risk: A review}
%\title{Network models of financial systemic risk: A survey
%\title{Network approaches to financial systemic risk: A survey

%\tk{Is there any better title? I left some ideas in text.}


\author{Fabio Caccioli,$^{1,2,3}$ Paolo Barucca,$^{4,5}$ and Teruyoshi Kobayashi$^{6}$\thanks{Corresponding author: kobayashi@econ.kobe-u.ac.jp}
}
\date{
$^1$\emph{Department of Computer Science, University College London, London, UK} \\
$^2$\emph{Systemic Risk Centre, London School of Economics and Political Science, London, UK} \\
$^3$\emph{London Mathematical Laboratory, London, UK}\\
$^4$\emph{Department of Banking and Finance, University of Zurich, Zurich, Switzerland}\\
$^5$\emph{London Institute for Mathematical Sciences, London, UK} \\
$^6$\emph{Graduate School of Economics, Kobe University, Kobe, Japan} \\  [2ex]
\today
} 
 \maketitle


%{Fabio Caccioli \at  Department of Computer Science, University College London, London, UK\\ Systemic Risk Centre, London School of Economics and Political Science, London, UK\\
%London Mathematical Laboratory, London, UK
%\and Paolo Barucca \at FINEXUS Center, University of Zurich, Zurich, Switzerland
%\and $^*$Corresponding author, Teruyoshi Kobayashi \at
%              Graduate School of Economics, Kobe University, Kobe, Japan \\
%              \email{kobayashi@econ.kobe-u.ac.jp}                    
%}

%\date{\today}

\begin{abstract}%150--250 workds %currently 170 words
The global financial system can be represented as a large complex network in which banks, hedge funds and other financial institutions are interconnected to each other through visible and invisible financial linkages. 
Recently, a lot of attention has been paid to the understanding of the mechanisms that can lead to a breakdown of this network. This can happen when the existing financial links turn from being a means of risk diversification to channels for the propagation of risk across financial institutions. 
In this review article, we summarize recent developments in the modeling of financial systemic risk. We focus in particular on network approaches, such as models of default cascades due to bilateral exposures or to overlapping portfolios, and we also report on recent findings on the empirical structure of interbank networks. 
The current review provides a landscape of the newly arising interdisciplinary field lying at the intersection of several disciplines, such as network science, physics, engineering, economics, and ecology.
\end{abstract}

%\keywords{Financial networks \and systemic risk \and contagion \and clearing algorithm \and overlapping portfolio \and core-periphery
%\and Overlapping portfolio Credit Risk \and Default \and Clearing Systems \and Payment system \and Central bank \and interbank markets
%PACS: 89.65.Gh
%}% \tk{choose 4--6 keywords}
%JEL classification: D62; E42: E44:E58 
%\keywords{Financial networks \and Systemic risk \and Interbank markets \and Overlapping portfolio Credit Risk \and Default \and Clearing Systems \and Payment system \and Systemic risk \and Central bank}

\section{Introduction}
\label{intro}

 Since the global financial crisis of 2008--2009, many studies on financial systemic risk have been accumulated to date. One of the most distinctive features of this newly arising field is its interdisciplinary nature, with researchers having backgrounds in economics and finance, statistical physics, ecology, engineering, applied mathematics, etc~\cite{May2008Nature,Schweitzer2009Science,Battiston2016Science}. The study of financial systemic risk has attracted such a diversity of disciplines because financial markets are complex systems, and the study of complex systems has traditionally been an interdisciplinary field. 
 
%  \del{Studying financial systemic risk is all about understanding the complexity of the network of financial linkages formed by various sorts of financial transactions~\cite{Haldane2011systemic,Fouque2013handbook}.}\fc{The financial network is important, but I do not think it is only about it. I would smooth a bit the sentence. I propose something like the paragraph below (though I am not yet 100\% satisfied with it)}

 In the financial market there is a wide variety of market participants, such as commercial banks, insurance companies, hedge funds, individual investors, and central banks. These participants are interacting with each other by selling and buying financial assets, creating complex webs of financial liabilities, cross-asset holdings, and correlations in asset returns.
 Financial systemic risk is, loosely speaking, the risk associated with the occurrence of a breakdown of the financial system. Looking at systemic risk from the point of view of complex systems means thinking of it as emerging from the interactions between different players that operate in the financial market. Moreover, individual participants react to the aggregate dynamics of the market that they collectively create, so that to understand systemic risk the feedback loop between individual and collective dynamics has to be accounted for.
 A key to understanding systemic risk is thus to uncover the mechanisms that lie behind the micro-macro feedback.
 Because of the fact that many interactions that take place in financial markets can be represented as a network of financial linkages between institutions, a significant fraction of research in systemic risk has been devoted to the study of financial networks ~\cite{haldane2011systemic,Fouque2013handbook}, which is the focus of this short review.

      In this article, we provide a review of recent studies on financial systemic risk. Here, we focus on network models of financial markets that has been developed outside traditional economics, although there are also many studies of systemic risk in the literature of economics whose approaches are typically based on game theory, finance and macroeconomic modeling. An advantage of modeling the financial system as a complex network is that we can directly analyze complex feedback between micro- and macroscopic phenomena without oversimplifying the structure of financial linkages. It is the structure of networks that plays an essential role in leading micro events to collective phenomena. Moreover, the empirical structure of financial networks rarely takes a stylized form such as the \ER random graph and a star graph, but it takes more complex structures such as multiplex, bipartite, core-periphery, and time-varying networks, depending on the property of the financial linkage concerned. Such complex yet realistic network structures cannot be treated in the traditional economic models.

 In section~\ref{sec:clearing}, we first explain basic clearing algorithms that are needed to compute the allocation of debtor's assets among creditors. In normal times, a bilateral credit contract is settled when a debtor repays in full the amount of borrowed funds. However, if a debtor fails, then it is no longer straightforward to know how to consistently allocate the debtor's remaining assets across its creditors. It may seem reasonable to allocate the remaining assets proportionally to the amount of funds lent, but the losses that the creditors incur may also cause their defaults.   
 If such contagious defaults happen, the remaining assets of the failed creditors should also be allocated proportionally to their creditors, which might in turn lead to a collapse of the creditors' creditors, and so on. Therefore, in the presence of contagious default cascades, calculating the final allocation of funds is essentially equivalent to computing a fixed point of an iterative map. 
 We provide a brief introduction to a widely known clearing algorithm, the Eisenberg-Noe algorithm~\cite{eisenberg2001systemic}.

 The allocation of funds achieved by the Eisenberg-Noe' clearing algorithm is economically sensible in the sense that the amounts of funds paid back to the creditors are endogenously determined in proportion to the amounts of funds lent. In the real world, however, such an ideal allocation may no be feasible, especially in financial turmoil, because evaluating the assets of failed banks and negotiating among creditors takes a long time. Therefore, many studies consider a fixed exogenous recovery rate (i.e. creditors receive a fixed percentage of their payment), which is often set to zero to analyze worst case scenarios. An advantage of imposing the zero-recovery assumption is that it allows us to analyze financial contagion by using a model of social contagion that has been developed in network science. Pioneers of this approach are Gai and Kapadia~\cite{gai2010contagion}, who exploited the social cascade model of Watts~\cite{Watts2002}.    
  We review some of the models of this type of default cascades in section~\ref{sec:interbank_cascades}.
  
  Propagation of distress between borrowers and creditors can also occur before the borrower's default, because of credit quality deterioration. In section \ref{sec:distress_propagation}, we review a model, the so-called DebtRank \cite{battiston2012debtrank}, which has been introduced to account for this situations, together with some of its extensions.
 
  Stress does not only propagate between borrowers and lenders. In fact, losses can also propagate between investors having common assets. For example, the devaluation of an asset that is commonly held by many banks would simultaneously undermine the balance sheets of the banks holding the asset. If the loss of a bank is so severe that the bank is unable to meet the minimum requirement for its capital ratio, then the bank will have to sell some of its assets. This liquidation has a negative impact on the prices of the assets that are being sold, which causes losses to other banks holding these assets. In this way, a cascade of defaults could be triggered by the initial decline in the price of an asset, and fueled by the presence of overlapping portfolios among banks. Modeling the hidden interbank linkages formed through cross asset holdings is then important to understand systemic risk. Recent studies in this domain are introduced in sections~\ref{sec:overlapping}. 
  
In section~\ref{sec:structure}, we summarize recent work on the structure of empirical interbank networks and its dynamics.
Whether the initial default of a bank can trigger a large-scale default contagion depends largely on the way financial institutions are connected to each other. We introduce studies that examined the empirical structure of networks formed by interbank bilateral trades and explain the well-studied core-periphery structure and its possible limitations. The necessity to analyze daily network dynamics is also discussed.
  Section~\ref{sec:discussion} concludes.

 \section{Clearing algorithms}\label{sec:clearing}

% Financial institutions are entangled in networks of complex contracts, involving two or more counterparties at the same time. Contracts may vary in terms of complexity, due to interdependencies, number of parties involved, seniority, maturity, collateralization, and each financial instrument needs a specific mathematical formulation.
% The first structural models of systemic risk focused on simple financial contracts with elementary mechanical rules for clearing, i.e. the set of financial operations leading a commitment to its settlement. 
% In order to provide analytic insights, literature on clearing algorithms of financial networks has mainly focused on the modeling of cross-holdings, simple debt contracts, involving two parties, and simple derivatives - such as Credit Default Swaps (CDS). 
 
In the context of financial networks and systemic risk, clearing plays a central role as the settlements of different transactions are entangled in a network of mutual commitments. Hence, it constitutes the structural basis leading to observable liquidity problems, failed payments, losses, and insolvencies. In full generality, financial transactions can be over-the-counter (OTC) contracts, where each pair of counterparties has to settle its own contract, or can be collected and managed by Central-Clearing-Counterparties (CCPs) - absorbing all or some of the financial risks - or be mediated by a market via an order book dynamics.

Here, we introduce a well known clearing model, the one proposed by Eisenberg and Noe \cite{eisenberg2001systemic}, which led to a series of works on the problem of valuating systemic risk in financial networks.
In a schematic representation, a debt contract between two counterparties is represented by an amount $L_{ab}$ to be paid at time $T$ by the financial institution $A$ to financial institution $B$. Though this contract is simple in nature, it is easy to realize the complexity that may arise by thinking that, in the case of conflicting financial obligations, institution $A$ might be unable, or even unwilling, to repay $B$ in full at the prescribed time $T$. 
It is therefore necessary to consider the seniority of the debt, i.e., the priority of its repayment with respect to other obligations. Besides deterministic quantities, the counterparty $B$ is interested in knowing ex-ante - before the maturity of the contract - the probability of default (PD) of $A$, i.e., the probability of $A$ not honoring the contract, and the loss given default (LGD), that is the credit that $B$ will be able to recover given the default event. 

In this section, we focus on a deterministic clearing procedure \cite{eisenberg2001systemic} and in particular we will investigate a specific case of an arbitrary number of financial institutions entangled in a mono-layer network of simple bilateral debt contracts. 
All institutions in the model need to clear their payments at the time of maturity, same for all bilateral contracts. Eisenberg and Noe provide a solid set of assumptions to study the basic properties of the solution of this kind of clearing procedure. This methodology allows to compute cascades of defaults, reallocation of funds, and systemic effects when dealing with network of contracts. 
It is important to stress that these models constitute only one mathematical aspect of the complex and multidisciplinary problem of ensuring financial stability, involving political, legal, economic, financial, and infrastructural aspects. Opacity and information asymmetry in markets cause a general discrepancy between credit risk estimations by different financial institutions and the real risks. It is arguable, given the complexity of system, that even a fully-fledged structural model would not be able to correctly compute credit risks. As a consequence, opacity and financial complexity lead to a systematic inefficiency in risk taking, both in the construction of portfolios and in the issuing of credit \cite{Battiston2016PNAS}.

\subsection{The Eisenberg-Noe model}

Following closely the work by Eisenberg and Noe \cite{eisenberg2001systemic}, let us consider an economy composed of $N$ financial institutions, hereafter called banks for the sake of simplicity. Each bank has nominal liabilities to other banks that need to be settled at the same time. 
Such structure of liabilities can be represented with a $N\times N$ matrix of non-negative real numbers $L$, where each entry $L_{ij}$ stands for the nominal liability of node $i$ to node $j$. 
Nominal liabilities are all non-negative because a debt contract $L_{ij}$ between bank $i$ and bank $j$ with a negative value would constitute an effective credit for bank $i$ and therefore a debt for bank $j$, and would simply appear as a positive value of the entry $L_{ji}$.
A further realistic assumption is the absence of nominal claims of a bank against itself, which results in having null elements on the diagonal of the liabilities matrix.  

Finally, each bank has a non-negative operating cash flow, $e_i$, representing the net cash received by each bank from the outside of the financial system under consideration.
External liabilities can be introduced either by setting a negative value to cash flows or by adding an extra bank in the liabilities matrix. Such fictitious bank has zero cash flow, i.e. $e_0 = 0$, and is supposed to receive an amount $L_{i0}$ from each financial institution $i$. Nevertheless, this choice is not entirely equivalent to having a negative $e_i$, as the seniority - the priority that a contract takes - of the liabilities in the matrix $L$, including of the ones towards node $0$, is lower than the one that would result from simply subtracting $L_{i0}$ from the non-negative cash flow $e_i$. In the following we will take the latter choice and consider the possibility of an extra bank.

In this simplified picture, a financial system $\mathcal{F}$ is a pair of a non-negative liabilities matrix and a operating cash flow vector, $\mathcal{F}=(L,\,e)$.
This framework excludes the existence of many realistic characteristics of financial systems, such as: multiple contracts between a given pair of banks, multiple levels of seniority, or involving more than two banks, or also, different times to maturity. Further, it does not model in detail the stochastic features of cash flows, nor their correlation structure or the existence of common asset holdings among different institutions.

Despite its specificity, this framework successfully deals with the problem of identifying a clearing vector of payments, i.e. a vector that associates to each bank the total amount that it is able to repay to its debtors given the financial system, $(L,\,e)$, and, in doing so, it is able to introduce some crucial quantities and features of systemic events in financial systems, such as the dynamics of the contagion process and the conditions for the existence of unique solutions.

Let us now enter in more detail into the definition of the clearing procedure as defined in \cite{eisenberg2001systemic}. It is useful to introduce a few auxiliary variables, i.e. the total nominal obligations $\bar{p}_i$ defined as: 
\begin{align}
\bar{p}_i = \sum_{j=0}^NL_{ij},\label{eq:pbar}
\end{align}
and the relative liabilities matrix, quantifying the fraction of liabilities from bank $i$ that a bank $j$ is entitled to receive in the case of full repayment,
\begin{align}
\Pi_{ij}=
\begin{cases}
\frac{L_{ij}}{\bar{p}_i},\; \text{if}\;\bar{p}_i >0\\
0\;\text{otherwise}.
\end{cases}
\end{align}
Finally, we define the payment vector $p$, the vector of unknown quantities that we want to identify, that is the amount that each bank is actually able to repay. By definition, all the elements of the payment vector are less than or equal to the elements of the obligation vector $\bar{p}$ and greater or equal to zero, or in formulas, $p\leq \bar{p}, \; p\geq 0$.
The clearing procedure may yield multiple solutions for $p$, but, by making some intuitive and natural financial assumptions, it can be shown that the solution is unique.

The financial requirements for the Eisenberg-Noe clearing procedure are the following: (i) all the elements of the payment vector are less than or equal to the available cash flow of the bank (Limited Liabilities), (ii) banks repay as much as they can, i.e. they are not allowed to keep cash in their balance as long as they have not fully repaid all their liabilities; this can also be expressed that the equity has a lower seniority with respect to interbank liabilities; (Absolute Priority) (iii) the individual payment of a given liability, i.e. the effective value repaid to a bank, has to be proportional to the fraction of the total obligation that the liability represents, as given by the relative liabilities matrix $\Pi_{ij}$ (Proportionality). More explicitly, for each institution the ratio between the liability repaid to a given counterparty and the total amount repaid to all counterparties has to equate the ratio between the nominal liability and the total amount of liabilities that the institution has. 

From these simple assumptions, we can compute the net position of a bank assuming a given payment vector $p$. In fact, the total assets of bank $i$ will amount to $e_i +\sum_j \Pi_{ji}p_j$, whilst having a total obligation in the interbank market $\bar{p}_i$. Given assumption (ii), as long as the obligation is less than the total amount of assets then the bank will repay in full, i.e. $p_i=\bar{p}_i$, and will remain with a net position $e_i +\sum_j \Pi_{ji}p_j - \bar{p}_i$. Otherwise, when $e_i +\sum_j \Pi_{ji}p_j < \bar{p}_i$, bank $i$ will have to use all its assets to repay its counterparties, i.e. $p_i=e_i +\sum_j \Pi_{ji}p_j$, leaving the bank with a net position equal to zero and a total amount of unrepaid debt equal to $\bar{p}_i - e_i -\sum_j \Pi_{ji}p_j$. 
Hence, if we require all banks to simultaneously satisfy the same relations, we derive the following system of equations:
\begin{align} 
p_i = \min\left\{e_i+\sum_{j}\Pi_{ji}p_j,\,\bar{p}_i\right\}\,\;\forall\: i=1,\ldots , N, \label{eq:pay_vec}
\end{align}
which defines a fixed-point problem for the payment vector, whose components also have to satisfy the conditions $\bar{p}_i\geq p_i \geq 0$ for each bank $i$. 
\subsection{Convergence to the payment vector}
The system, \eqref{eq:pay_vec} yielding the solutions to the clearing procedure, is a set of non-linear equations, for which the dependence on each component of the payment vector is piece-wise linear, monotone, bounded, and continuous. All these properties allow to demonstrate the existence and uniqueness of the solution, by means of the general Knaster-Tarski theorem~\cite{tarski1955lattice}.  

To identify the solution, Eisenberg and Noe propose a simple iterative algorithm, where each iteration is made of two-steps: first, an update that identifies the set of defaulted banks
\begin{align} 
D(\mathbf{p}) = \{i\in \{1,\,...N\}\, | p_i < \bar{p}_i\}.\label{eq:default}
\end{align} 
Secondly, the payment vector is updated by looking for the fixed-point of the following map: 
\begin{align} 
\vec{p'} = \Lambda(\vec{p'})(\Pi(\Lambda(\vec{p'})\vec{p} + (1-\Lambda(\vec{p'}))\bar{\vec{p}} + \vec{e}) + (1-\Lambda(\vec{p'}))\bar{\vec{p}},\label{eq:FF}
\end{align} 
where $\Lambda(\vec{p'})$ is a diagonal matrix, such that $\Lambda(\vec{p'})_{ii}$ is one if $i\in D(\vec{p'})$, and zero otherwise. Equation \eqref{eq:FF} admits a unique fixed-point if the financial network is regular; the regularity condition is defined in full detail in \cite{eisenberg2001systemic}, but for the sake of brevity we just notice that such condition is easily respected when each bank in the system has a strictly positive equity. 

The fixed-point $\vec{p^*}$ of the vector equation \eqref{eq:FF} then is used to update the set of defaulted banks $D$ accordingly. 
The two-steps are repeated until convergence to a set of defaulted banks $D$ and payment vector $p$ satisfying \eqref{eq:pay_vec}, and under the regularity condition such payment vector is the only solution to \eqref{eq:pay_vec}.

\subsection{Related literature and generalizations}
The seminal paper by Eisenberg and Noe (EN, hereafter) ignited research and many generalizations have been proposed since its publication. Here we discuss only the ones that introduce the most important effects and outline their consequences on the properties of the systemic losses. 

Rogers and Veraart \cite{Rogers2013} introduce default costs in the system. In fact, when insolvency occurs it is unrealistic to assume - as EN do in their clearing model - the absence of additional costs: an insolvent financial institution may (i) need to rapidly liquidate valuable external assets at a lower price than its present market valuation, e.g. due to the price impact of a fire-sale, and (ii) need to withdraw their interbank assets conceding their counterparty a discount related to the early repayment before maturity. With such motivation the EN system is modified, introducing two factors $\alpha$ and $\beta$ - both between zero and one -, that discount respectively the external, and the interbank assets of an insolvent institution. The resulting system of equations reads: 
\begin{align}
p_i=
\begin{cases}
\bar{p}_i,\; \text{if}\;\bar{p}_i < e_i + \sum_{j}\Pi_{ji}p_j,\\
\alpha e_i + \beta \sum_{j}\Pi_{ij}p_j\;\text{otherwise}.
\end{cases}\label{eq:rogersveraart}
\end{align}
It is easy to recognize that when $\alpha=\beta=1$ the system of equations \eqref{eq:rogersveraart} is equivalent to \eqref{eq:pay_vec}. Otherwise, these discount factors increase losses, introducing costs for bailouts that exceed the initial losses of the financial system. In fact, while EN simply redistributes losses across the financial system, as discussed in \cite{visentin2016rethinking}, Rogers and Veraart clearing procedure recognizes the existence of extra costs, which are exactly the ones that financial regulatory institutions want to minimize. The motivation is that, while external losses coming from economic shocks are driven by scarcely controllable complex economic dynamics, the possible endogenous loss amplification due to the financial system interconnectedness could be avoided by regulators by imposing specific policies. 

Both the EN and the Rogers-Veraart models are completely deterministic. They constitute a mechanism, respectively, for the redistribution and amplification of losses
in a financial system. Nevertheless, such mechanism is triggered only by an actual insolvency, i.e. the liabilities have to exceed assets for a bank to propagate its losses. They are default contagion mechanisms. This is a limitation of the clearing framework that could or may be overcome in two ways: first, directly account for a propagation that activates when a counterparty is in financial distress, as in DebtRank \cite{battiston2012debtrank}, and define a distress contagion mechanism; secondly, account - before maturity - for the uncertainty on the value that the external assets will take at maturity, i.e. introduce a probability distribution over external assets. In the latter, the propagation remains based on a default contagion mechanism. 

Nevertheless, the distribution on the external assets will include scenarios that on average will cause expected losses, that would have been absent, if the external assets were taken at their present value before maturity. This is the approach taken by Elsinger et al.~\cite{Elsinger2006}. A recent effort was made to consider the two kinds of mechanisms in a unified framework of network valuation \cite{barucca2016network} accounting for uncertainty, and default costs in a compact form.

\section{Cascades of bank defaults due to bilateral interbank exposures}\label{sec:interbank_cascades}

\subsection{The Gai-Kapadia model}
 Since the work of Gai and Kapadia~\cite{gai2010contagion}, many researchers have developed network models of default cascades in financial networks, especially interbank networks in which banks lend to and borrow from each other. The basic structure of the Gai-Kapadia cascade model is heavily based on the Watts model of global cascades~\cite{Watts2002} that would occur on networks formed by social interactions between humans.
 In the Watts  model, the mechanism of how a node affects its neighbors is quite simple; a node gets ``activated'' (or ``infected") if and only if at least a certain fraction $R\in[0,1]$ of its neighbors are activated. The Watts model of cascades is therefore categorized as a \emph{linear threshold model} or just a \emph{threshold model}, which belongs to the class of \emph{complex contagion}.\footnote{In contrast, \emph{simple contagion} refers to a contagion process in which the probability of a node being affected by its neighbor is exogenously given.}  The main implication of the Watts model is that, on randomly connected networks, even a vanishingly small fraction of initial active nodes may lead a significant fraction of infinitely many nodes to get activated as long as the network is not too sparse or too dense. This critical phenomenon is called a \emph{global cascade}, and the analytic condition under which a global cascade may occur, called the \emph{cascade condition}, can be computed by exploiting a mean-field approximation. 
 
  In this section, we fist explain the basic properties of the threshold model developed by Watts~\cite{Watts2002}. Since the threshold of activation for humans can be reinterpreted as the threshold of defaults for banks, understanding the threshold model used in the social network literature is important to understand many existing models of default contagion in interbank networks. Here, we explain two different approaches to calculating the size of cascades, namely the tree-based approximation~\cite{Gleeson2007} and the generating function approach~\cite{gai2010contagion,Watts2002}.
  While the Watts model assumes that edges are undirected, extending its framework to directed networks is straightforward~\cite{gai2010contagion}.

 \subsubsection{Tree-based approximation}
  Here, we briefly explain a tree-like approximation method for solving the model of social contagion~\cite{Gleeson2007}. The basic idea of a tree-like approximation is to calculate the average final fraction $\rho$ of activated nodes by assuming that the network is locally tree-like. The probability of a randomly chosen node being active, or the average size of global cascades, is calculated by the following equation:
  \begin{align} 
 \rho = \rho_0 + (1-\rho_0)\sum_{k=1}^{\infty}p_k \sum_{m=0}^k \begin{pmatrix} k \\ m \end{pmatrix} q^m (1-q)^{k-m}F\left( \frac{m}{k} \right),\label{eq:rho}
 \end{align}
 where $q$ is the probability that a randomly chosen neighbor is active, $\rho_0$ is the chance that a node is initially active (i.e., a seed node), $p_k$ is the degree distribution, and $k$ and $m$ stand for the degree and the number of active neighbors, respectively. At this point, the neighbors' activation probabilities, which are considered to be identical to the mean value, are regarded as independent since  there is no cycle of influence at least locally thanks to the assumption of a locally tree-like structure.   
 In the simple Watts model, the response function $F(\cdot)$ takes 1 if $m/k > R$, and 0 otherwise. 

    \begin{figure}[t]
\begin{center}
           \includegraphics[width=.55\columnwidth]{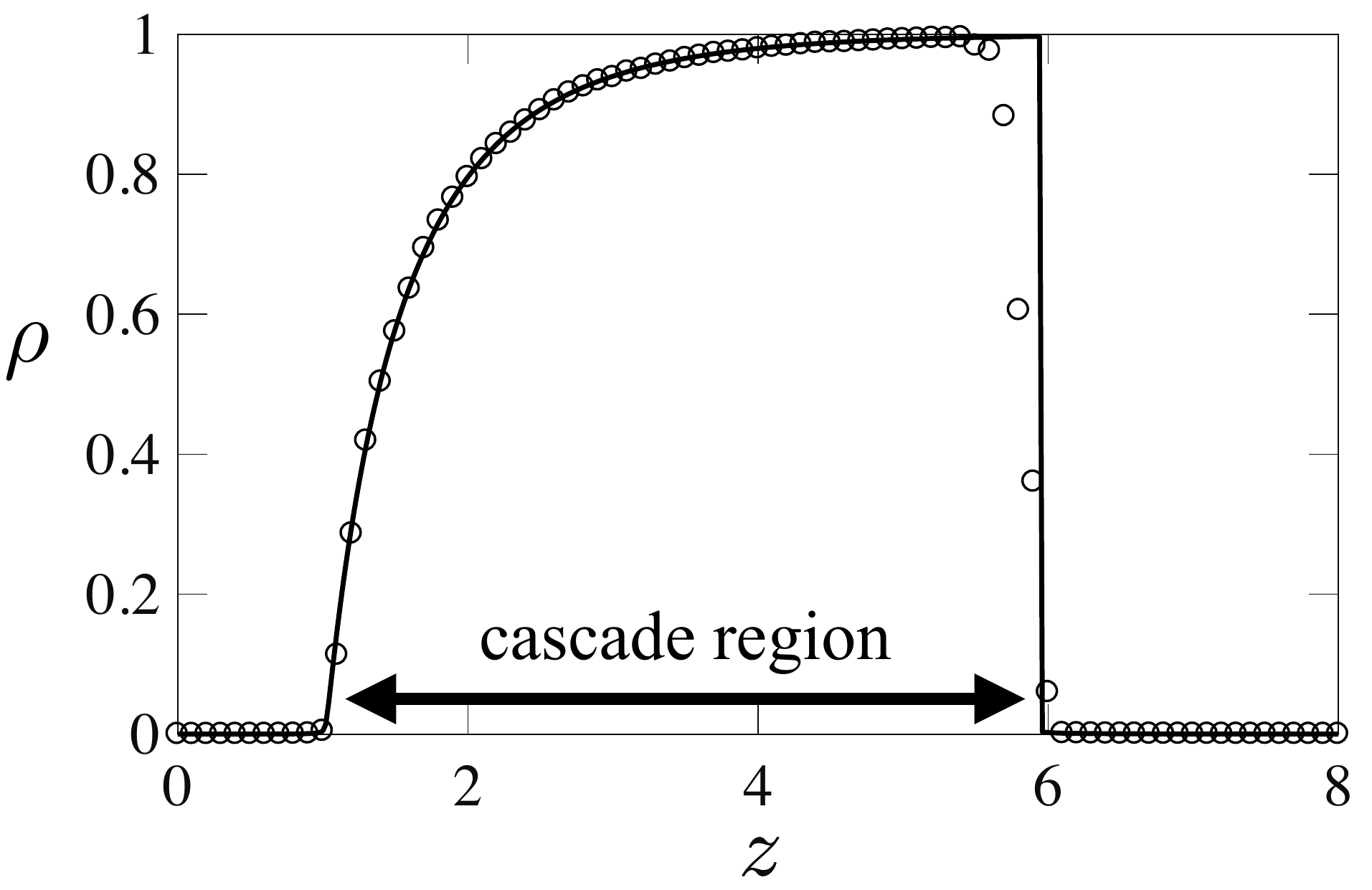}
                \end{center}
    \caption{Threshold cascade model with a Poissonian degree distribution. Line denotes the value of $\rho$ calculated from Eq.~\eqref{eq:rho}, and circle represents the simulated average cascade size, averaged over 1000 runs with $N=10^{5}$ and $R=0.18$. $z$ is the mean degree, and seed fraction is $\rho_{0}=10^{-4}$.}
 \label{fig:threshold_model}
\end{figure}
 
 The probability $q$ is given as
  \begin{align} 
 q = \rho_0 + (1-\rho_0)\sum_{k=1}^{\infty}\frac{k}{z}p_k \sum_{m=0}^{k-1} \begin{pmatrix} k-1 \\ m \end{pmatrix} q^m (1-q)^{k-1-m}F\left( \frac{m}{k} \right), \label{eq:recursion_q}
 \end{align}
 where $z$ denotes the mean degree, which is the connectivity parameter of the network. Note that one should use the excess degree distribution  $kp_{k}/z$, as opposed to the degree distribution $p_{k}$, to compute the average fraction of active neighbors of a neighbor. 
 This can be understood as a situation in which a ``child node" is influenced by its ``parent nodes" and then the child node affects the ``grandchildren", and so forth.
 The solution for $q$ is then obtained as a fixed point of recursion equation \eqref{eq:recursion_q}, which in turn gives the solution for the mean cascade size $\rho$ from Eq.~\eqref{eq:rho}.\footnote{When there are multiple fixed points, the smallest solution is selected as the valid solution. In fact, depending on the model parameters, recursion equation \eqref{eq:recursion_q} may exhibit a saddle-node bifurcation, which could cause a phase transition~\cite{Gleeson2007}. Such a phase transition can be observed in a variety of threshold models~\cite{Lee2014PRE,KobayashiPRE2015}.}  
 Regardless of its simplicity, the tree-based method can predict the final size of a global cascade very accurately (Fig.~\ref{fig:threshold_model}).
 
 The parameter space within which a global cascade may occur is called the cascade region, and the analytical condition for the parameters to be satisfied in the cascade region is called the cascade condition. To see the derivation of the cascade condition, let us define $S(q)$ as
 \begin{align}
     S(q)\equiv \sum_{k=1}^{\infty}\frac{k}{z}p_k \sum_{m=0}^{k-1} \begin{pmatrix} k-1 \\ m \end{pmatrix} q^m (1-q)^{k-1-m}F\left( \frac{m}{k} \right). \label{eq:s_q}
 \end{align}
 Gleeson and Cahalane~\cite{Gleeson2007} argue that if the derivative of the RHS of Eq.~\eqref{eq:recursion_q} (i.e., $\rho_{0}+(1-\rho_0)S(q)$) near $q=0$ takes a value larger than one, then a vanishingly small initial seed $\rho_0$ results in a large value of $\rho$. The \emph{first-order cascade condition} is therefore given by
 \begin{align}
     (1-\rho_0)\sum_{k=1}^{\infty}\frac{k(k-1)}{z}p_k \left[ F\left( \frac{1}{k}\right)-F(0)\right] > 1.
     \label{eq:cascadecondition_first}
 \end{align}
 In fact, a comparison with simulation results reveals that this is not a very accurate condition for the parameter space $(R,z)$. Therefore, they also propose the \emph{second-order cascade condition} given by
 \begin{align}
     (C_1-1)^2 - 4C_{0}C_{2} + 2\rho_{0}(C_{1}-C_{1}^{2}-2C_{2}+4C_{0}C_{2}) < 0, \label{eq:cascadecondition_second}
 \end{align}
 where it is assumed that $\rho_0^2\approx 0$, and $C_\ell$ is defined such that $S(q) = \sum_{\ell=0}^\infty C_{\ell}q^\ell$ and
 \begin{align}
     C_\ell \equiv \sum_{k=\ell+1}^{\infty}\sum_{n=0}^{\ell}\begin{pmatrix} k-1 \\ \ell \end{pmatrix}\begin{pmatrix} \ell \\ n \end{pmatrix} (-1)^{\ell-n}\frac{k}{z}p_{k}F\left(\frac{n}{k}\right).
 \end{align}
 Condition \eqref{eq:cascadecondition_second} states that the second-order approximation of Eq.~\eqref{eq:recursion_q}, $q = \rho_0 +(1-\rho_0)(C_0 +C_1q+C_2q^2)$, has no solution near $q=0$, because the existence of a positive root near $q=0$ implies that a global cascade is impossible.\footnote{To obtain the second-order condition, we can simply approximate $S(q)$ near $q=0$ up to the second order: $S(q) = S(0) + S^{\prime}(0)q + \frac{1}{2}S^{\prime\prime} (0)q^2$, where $C_0 = S(0)$, $C_1 = S^\prime(0)$, and $C_2 = \frac{1}{2}S^{\prime\prime}(0)$. }
 Gleeson and Cahalane~\cite{Gleeson2007} showed that the second-order cascade condition \eqref{eq:cascadecondition_second} well matches the cascade region predicted by the numerical simulation.

\subsubsection{Generating function approach}
 
  Now we explain the generating function approach to calculating the expected cascade size. In doing so, we compute the probability that a randomly chosen node is \emph{vulnerable}; we say a node is vulnerable if its degree, $k$, satisfies $R\leq 1/k$. That is, if a node is vulnerable, then the node will get activated if at least one neighbor is active. Let $\mu_k = P[R\leq 1/k]$ denote the probability of a node having $k$ edges being vulnerable. Since the probability that a randomly chosen node has degree $k$ is $p_k$, the generating function of vulnerable node degree is given as
  \begin{align}
      G_0(x) = \sum_{k=0}^{\infty}\mu_{k}p_{k}x^k.
  \end{align}
Generating function $G_0(x)$ has information on all of the moments of the degree distribution only of vulnerable nodes.\footnote{See Ch.\ 13 of Newman~\cite{Newman2010book} for the basic explanation of generating functions.} The generating function for the excess degree distribution for the vulnerable nodes leads to
\begin{align}
    G_1(x) &=  \frac{\sum_{k=1}^{\infty}k\mu_{k}p_{k}x^{k-1}}{\sum_{k=1}kp_k} = \frac{G_0^{\prime}(x)}{z},
\end{align}
where $G^\prime$ represents derivative. Note that $G_1(1)$ is equal to the probability that a randomly chosen neighbor is vulnerable.

 We now introduce the generating function for the vulnerable cluster size:
 \begin{align}
     H_0(x) &= \sum_{n=0}^{\infty}\theta_nx^{n}, \\
     H_1(x) &= \sum_{n=0}^{\infty}\tilde\theta_nx^{n}, 
 \end{align}
where $\theta_n$ denotes the probability that a randomly chosen node belongs to a vulnerable cluster of size $n$, and $\tilde{\theta}_n$ is the corresponding probability for a neighbor of a randomly chosen node.
The generating function for the probability that a randomly selected neighbor belongs to a vulnerable cluster should satisfy the following self-consistency equation~\cite{Watts2002,Callaway2000PhysRevLett}: 
 \begin{align}
     H_1(x)  = 1 - G_1(1) + xG_1(H_1(x)). \label{eq:H1}
 \end{align}
The first term represents the probability that a neighbor is not vulnerable, and the second term corresponds to the size distribution of vulnerable clusters to which a chosen neighbor belongs. Note that if a node belongs to a vulnerable cluster of size $n$, then its neighbors must also belong to a vulnerable cluster of size $n$. Therefore, the generating function for the probability that a randomly chosen neighbor belongs to a vulnerable cluster of size $n$ depends on the second- and higher-order neighbor's generating function, resulting in a self-consistent determination~\eqref{eq:H1}~\cite{Watts2002,Callaway2000PhysRevLett}. Here, the assumption of a locally tree-like structure is needed to obtain the generating function, in which case different neighbors belong to (locally) independent subsets of a vulnerable cluster. 
Once $H_1(x)$ is obtained, $H_0(x)$ is computed as
\begin{align}
    H_0(x) = 1-G_0(1) + xG_0(H_1(x)), \label{eq:H0}
\end{align}
We note that, roughly speaking, the procedure for calculating $H_1$ ($H_0$) corresponds to the derivation of fixed point $q$ ($\rho$) in the recursion equation~\eqref{eq:recursion_q} (Eq.~\eqref{eq:rho}) in the Gleeson-Cahalane's~\cite{Gleeson2007} tree-based method.\footnote{This correspondence between the generating function approach and the tree-based method is not rigorous in the sense that the size of a vulnerable cluster is not identical to the average cascade size~\cite{Watts2002}.}

 The average vulnerable cluster size is given by 
 \begin{align}
 \langle n\rangle &= H_0^{\prime}(1) \notag \\
                  &= G_0(1) + \frac{(G_{0}^{\prime}(1))^2}{z - G_0^{\prime\prime}(1)}.
 \end{align}
It follows that the cascade condition is expressed as 
\begin{align}
    z < G_0^{\prime\prime}(1) = \sum_{k=1}k(k-1)\mu_{k}p_{k}. \label{eq:watts_cascade}
\end{align}
Note that since $\mu_k = F(1/k)$ for $k>0$, Eq.~\eqref{eq:watts_cascade} is equivalent to the previous cascade condition derived from the Gleeson-Cahalane's method (Eq.~\eqref{eq:cascadecondition_first}) as long as the threshold value is strictly positive (i.e., $R>0$ and $F(0)=0$) and the seed fraction is sufficiently small (i.e., $\rho_0\to 0$).

 \subsubsection{Financial contagion}
 
\begin{figure}[t]
\begin{center}
           \includegraphics[width=.55\columnwidth]{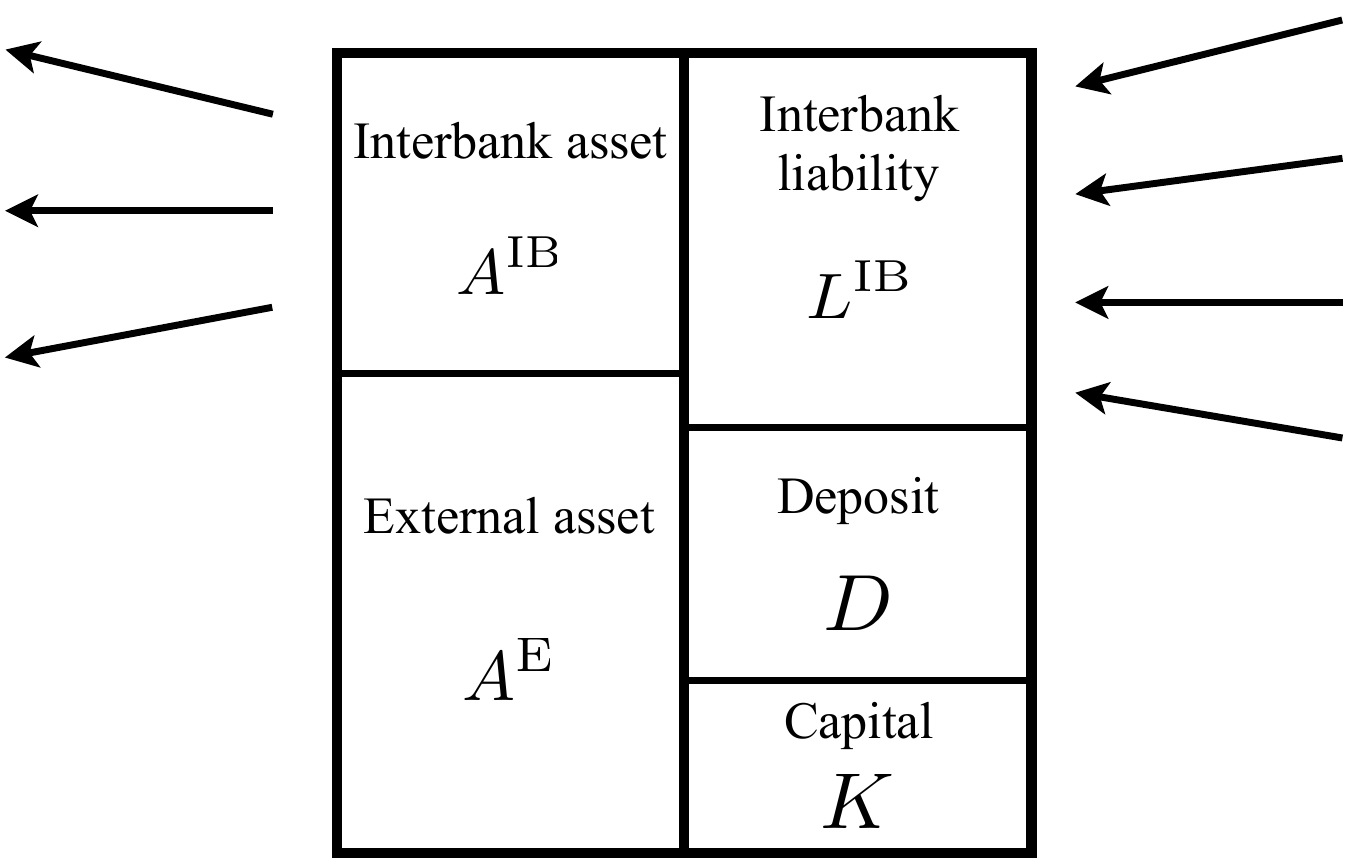}
                \end{center}
    \caption{Stylized balance sheet of a bank.}
 \label{fig:BS}
\end{figure}

  In models of financial contagion, nodes and directed edges represent banks and lending-borrowing relationships, respectively, and an activation of a node is interpreted as a bank default.
  In fact, the Gai-Kapadia model is isomorphic to the Watts model, the difference being that the former treats a directed random graph while the latter focuses on an undirected random graph. To see this, let us consider a stylized balance sheet of a bank (Fig.~\ref{fig:BS}). Suppose that each bank may have two types of assets: interbank assets, $A^{\rm IB}$, and external assets, $A^{\rm E}$ (such as stocks, bonds, etc). On the liability side, there can be interbank liabilities, $L^{\rm IB}$, and deposits from customers, $D$. Then, the solvency condition for bank $i$ is given by   
 \begin{align}
  A_{i}^{\rm IB} + A_{i}^{\rm E} - L_{i}^{\rm IB} - D_{i} > 0,
 \end{align} 
 which is equivalent to saying that the net worth (or the capital) of a bank should be positive.  
 
 Now, consider the situation in which the amounts of loans extended from a bank to other banks are evenly distributed, so that each interbank exposure is simply written as $A_{i}^{\rm IB}/|\mathcal{N}_{i}|$, where $\mathcal{N}_{i}$ denotes the set of borrowers to which bank $i$ lends. We assume that the ratio of total interbank assets, $A_{i}^{\rm IB}$, to net worth, $K_{i}$, is common across banks and is given by $A_{i}^{\rm IB}/K_{i} = 1/\overline{R}$ $\forall i$ for $\overline{R}>0$. Under these assumptions, the default condition for bank $i$ leads to
 \begin{align}
  \phi_{i} &> \frac{K_{i}}{A_{i}^{\rm IB}} = \overline{R}, \label{eq:threshold_condition}
 \end{align} 
  where $\phi_{i}$ is the fraction of bank $i$'s counterparties that have defaulted. The loss given default is assumed to be 100\% for simplicity; the lender would loose the full amount of funds lent to a defaulted bank. Eq.~\eqref{eq:threshold_condition} states that bank $i$ will default if the actual fraction of defaulted counterparties exceeds a constant threshold level $\overline{R}$, which is essentially the same mechanism as that of social contagion in the Watts model (i.e., just replacing $R$ with $\overline{R}$). 
    Note that if bank $i$ holds a sufficient amount of capital buffer such that $\overline{R}>1$, then there is no chance for bank $i$ to (contagious) defalt simply because the capital buffer can fully absorb the maximum possible losses.
  Since edges exposed to the risk of bank default are out-going edges (i.e., lending to other banks), the mechanism of social contagion applies here in a straightforward manner as long as there are no bidirectional edges.\footnote{The presence of bidirectional edges, although they rarely appear in empirical financial networks~\cite{Kobayashi2017arxiv}, could affect the size of contagion~\cite{Boguna2005PRE,Payne2011PRE}.}
  The trick here is that the volumes of interbank exposures, or edge weights, are evenly distributed between borrowers to which a bank lends. This greatly simplifies the analysis because otherwise we have to replace $\phi_{i}$ in the default condition \eqref{eq:threshold_condition} with the total fraction of losses that bank $i$ incurs.  A more general case of heterogeneous edge weights will be discussed in the following section.  
  
   Given the isomorphic property, it is straightforward to analyze cascades of bank defaults in the same way with the methods developed by Gleeson and Cahalane~\cite{Gleeson2007} and Watts~\cite{Watts2002}. Due to its simplicity and analytical tractability,  over the past years the Gai-Kapadia model has been frequently used as a baseline framework of more sophisticated models of financial contagion. We review some of these extensions in the next section.

\subsection{Extensions of the threshold cascade model}

  The Gai-Kapadia model spurred a flurry of research on financial contagion due to interbank exposures.  Before reviewing these works, let us summarize some of the important assumptions made in the simplest version of the Gai-Kapadia model; i) loans are evenly distributed, ii) interbank lending forms a \ER random graph, and iii) the risk of external assets is not considered. Recently, various models are proposed to make the cascade model more realistic by relaxing these assumptions. In considering the possible extensions of the threshold financial cascades, we can take advantage of the isomorphic property that characterizes the Watts and Gai-Kapadia models. That is, any models proposed as extensions of the Watts model could be applied to the model of financial contagion as well. 
  
 \subsubsection{Heterogeneous edge weights}
  
   When a heterogeneity of loan weights is introduced in the Gai-Kapadia model, the default condition is no longer captured by Eq.~\eqref{eq:threshold_condition}. The condition \eqref{eq:threshold_condition} is valid only if the amounts of interbank loans that a bank has lent to other banks are identical. Otherwise, the default condition cannot be expressed just by the fraction of defaulted borrowers, but rather expressed by the ratio of losses to total interbank assets:
    \begin{align}
  \frac{\sum_{j\in \mathcal{N}_i^{\rm def}}A_{ij}^{\rm IB}}{A_i^{\rm IB}} &> \overline{R}, \label{eq:non_identical_weight}
 \end{align} 
where $A_{ij}^{\rm IB}$ denotes the amount of funds lent from bank $i$ to bank $j$ (i.e., $A_i^{\rm IB} = \sum_{j}A_{ij}^{\rm IB}$), and $\mathcal{N}_i^{\rm def}$ is the set of bank $i$'s borrowers that have defaulted. Note that assuming identical edge weights, $A_{ij}^{\rm IB} = A_i^{\rm IB}/|\mathcal{N}_{i}|$, recovers condition \eqref{eq:threshold_condition} since the LHS would reduce to $|\mathcal{N}_{i}^{\rm def}|/|\mathcal{N}_{i}| \equiv \phi_i$. 

 Under a generic default condition \eqref{eq:non_identical_weight}, the standard mean-field approximation will not be appropriate since different borrowers have different weights, meaning that the number of defaulted banks itself is not informative. 
 One obvious way to analyze such a more general environment is to rely on numerical simulations.  
 Hurd and Gleeson~\cite{Hurd2013JCN}, Hurd~\cite{Hurd2016book} and Unicomb et al.~\cite{Unicomb2017threshold}, however, proposed alternative approximation methods to compute the solution of cascade dynamics.
   Hurd and Gleeson~\cite{Hurd2013JCN,Hurd2016book} consider a situation in which edge weights $w$ are random variables, whose CDFs $G_{kk^{\prime}}(w)$ depend on the degrees of nodes in both sides of an edge $k$ and $k^{\prime}$. They show that by imposing additional assumptions, one can obtain the analytical solutions for the mean cascade size.
   Unicomb et al.~\cite{Unicomb2017threshold} extend Gleeson's~\cite{GleesonPRL2011,GleesonPRX2013} approximate master equation, showing that an increased weight heterogeneity will reduce the size of cascades. 
  
\subsubsection{Non-\ER networks}
 
  The \ER random graph~\cite{Erdos1959PublMath} is probably the most widely used network structure in the analytical models of global cascades, yet no empirical financial networks exhibit the \ER structure. For example, it has been shown that the degree distribution of interbank networks follows fat-tail distributions such as a power-law distribution and a log-normal distribution~\cite{Boss2004,Iori2008JEDC,Cont2013}. Moreover, although the most analytical methods assume a locally tree-like structure, empirical networks have local clusters, and edges have a degree-degree correlation called assortativity~\cite{Newman2010book}.  
  
  It is natural to think that the tree-based method could be extended to the configuration graphs with arbitrary degree distribution $p_{k}$ as long as the network has a locally tree-like structure. A possible problem is that the tree-like assumption might no longer hold true once a more realistic network structure is considered. Even in the configuration model, for example, the clustering coefficient can be large when a fat-tail degree distribution is assumed~\cite{Newman2010book}.\footnote{In the configuration model, the clustering coefficient will approach 0 as the network size goes to infinity if the second moment of the degree distribution takes a finite value. See Eq.~(13.47) of \cite{Newman2010book} for details.} Fortunately, recent studies develop several ways in which the presence of local cycles would not affect the accuracy of the analytical solutions. Melnik et al.~\cite{MelnikPREunreasonable} provide some conditions under which the tree-like approximation works ``unreasonably" well even in networks with high levels of clustering. 
  Radicci and Castellano~\cite{RadicchiPREbeyond} develop an alternative technique based on a message-passing algorithm, which gives an accurate approximation in networks with local clusters.
    Ikeda et al.~\cite{Ikeda2010JP} also show that the presence of local clusters will enhance the chance of global cascades.

  Another possible departure from the \ER graph is that there are negative degree-degree correlations (or disassortativity) in real-world financial networks~\cite{Soramaki2007physicaA,BechPhysicaA2010}. That is, banks with high degrees are likely to trade with low-degree banks.
Dodds and Payne~\cite{Dodds2009PRE}, Payne et al.~\cite{Payne2009PRE,Payne2011PRE} and Hurd et al.~\cite{Hurd2017Plosone} study the effect of (dis)assortativity on the level of systemic risk, allowing for an arbitrary degree distribution. They show that assortativity of financial linkages strongly affects the expected cascade size.

 We can extend the tree-based method to even a multiplex structure of interbank networks. A multiplex network is a networked system consisting of multiple layers, on each of which nodes are connected by edges.\footnote{The essential difference between multiplex networks and multilayer networks is that in the latter, different layers can have different nodes while each layer has the same set of nodes in the former~\cite{Kivela2014_multilayer_review}.} 
 Interbank networks may exhibit a multiplex structure when banks trade different types of assets. For example, if there is a seniority in interbank assets (i.e., difference in the risk of loans), a ``monoplex'' model would no longer suffice. Brummitt and Kobayashi~\cite{Brummitt2015PRE} generalized the Gai-Kapadia model in a way that allows for different seniority levels, where different risk assets are traded in different layers. They consider a general case in which there are $M$ seniority levels, showing that the cascade condition is generally given by the trace of the Jacobian of $M$ many recursion equations. Of course, variety of interbank assets is not limited to seniority levels. There can be other ``layers'' in which banks trade long- and short-term assets, foreign exchange exposures, derivatives, and etc.~\cite{bargigli2015multiplex,Polenda2015JFS}. The multiplex structure of financial networks, however, is a relatively pre-matured research area in the sense that analytical models of financial contagion are still scarce.

\subsubsection{Risk of external assets}

  The only contagion channel considered in the Gai-Kapadia model is a cascade of repayment failures in the interbank market. In reality, however, this channel is just a part of the source of systemic risk. The risk of devaluation in external assets is one of the major concerns not only for the portfolio management of individual banks, but also for the systemic risk of the entire financial system. In recent years lots of works have been done on the contagion channel through overlapping portfolios, in which a fall in the price of an asset will  simultaneously affect many banks holding the same (or a correlated) asset~\cite{Beale2011,huang2013cascading,caccioli2014stability,caccioli2015overlapping}. Such a simultaneous shock to multiple banks has a potential to accelerate the traditional contagion process through interbank exposures. We will explain these studies in detail in sections~\ref{sec:distress_propagation} and \ref{sec:overlapping}. 
  
  In the Gai-Kapadia model, it is considered that the risk of external assets plays a role in initiating a contagion. Suppose that returns of external assets held by different banks are independent, and the price of an asset held by bank $i$ falls. If the devaluation of the asset is so large that the default condition \eqref{eq:threshold_condition} is satisfied (due to a reduction in $K_{i}$), then it may cause bank $i$'s creditors to default, initiating a contagion process. However, the role of external assets in reality is not that simple because the actual external assets are correlated, and the volatility of assets makes the health of balance sheets differ from bank to bank. To take into account these more realistic situations, many studies conduct simulations to understand the impact that a correlation in external assets has on systemic risk~\cite{Kobayashi2013EPJB}. Kobayashi~\cite{Kobayashi2014EL} provides a simple way to generalize response function $F$ to include the possibility that the value of external assets follow a probability distribution, which in fact corresponds to the Watts model in which the threshold for contagion is a random variable~\cite{Gleeson2007}.

 \section{Distress propagation due to credit quality deterioration}\label{sec:distress_propagation}

The importance of counterparty default contagion for practical purposes has been challenged both theoretically and empirically. From the theoretical point of view Glasserman et al.~\cite{glasserman2015likely} proved for instance that, within the Eisenberg-Noe framework, the contribution of contagion to the default probability of a bank is always small, and Battiston et al.~\cite{battiston2016rethinking} showed that this is the case because of a ``conservation of losses'' that is implicitly embedded in the Eisenberg-Noe algorithm, which prevents it from amplifying exogenous shocks.
From the empirical point of view, contagion analysis of real interbank systems have shown domino effects triggered by the failure of a small number of banks are unlikely to occur in practice \cite{upper2011simulation}.
On the other hand it was shown that networks of interbank exposures can significantly amplify distress propagation in presence of other contagion channels, such as for instance fire sales and overlapping portfolios \cite{caccioli2015overlapping}

Beyond its interaction with other contagion mechanisms, another reason why networks of interbank exposures can be important is the following: Models of contagion due to counterparty default risk assume that losses propagate from borrowers to lenders only after the default of a borrower. However, in practice, losses could occur even in absence of default, because of credit quality deterioration \cite{glasserman2015financial}. Consider the situation in which bank $i$ is exposed to bank $j$, which suffers a large loss. After the loss, the probability that $j$ defaults has increased, and therefore the expected cash flow associated with the exposures between $i$ and $j$ is reduced. If interbank assets were to be marked to market, this would mean that the value of the interbank asset of $i$ that is associated with its exposure to $j$ is reduced.
The idea of accounting for the propagation of distress before defaults led to the introduction of DebtRank~\cite{battiston2012debtrank}.

\subsection{DebtRank}
Let us consider a system of $N$ banks, and let us denote by $W_{ij}$ the interbank exposure of $i$ towards bank $j$, by $A_i^{\rm ext}$ the external (non interbank) assets of bank $i$, and by $L_i$ its total liabilities. DebtRank is a discrete-time map that describes the evolution of the equity of all banks after a shock hits the system. In the DebtRank dynamic banks can be in two states: active or inactive. An active bank is a bank that will pass distress to its creditors if subject to a loss, and a bank becomes inactive after it has passed distress to its creditors once. This does not necessarily mean that the bank has defaulted, nor that the bank cannot suffer additional losses, but simply that further losses will not be transmitted to its creditors.
If we denote by $h_i(t) = \frac{E_i(0)-E_i(t)}{E_i(0)}$ the relative loss of equity of bank $i$ at time $t$, and by $\mathcal{A}(t)$ the set of active banks at time $t$, the DebtRank dynamic reads
\begin{eqnarray}
h_i(t+1) &=& {\rm min} \left\{1, h_i(t) + \sum_{j\in\mathcal{A}(t)} \frac{W_{ij}}{E_i(0)} h_j(t) \right\}\\ \label{eq:debtRank}
%\mathcal{A}(t+1)&=&\{i~{\rm s.t.}~h_i(t)>0~{\rm and}~h_i(t-1)=0\}
\mathcal{A}(t+1)&=&\left\{i~|~h_i(t)>0,~h_i(t-1)=0\right\}
\end{eqnarray}
The meaning of the above dynamic is the following: the loss experience by bank $i$ between time $0$ and time $t+1$ is its loss up to time $t$ plus the new losses that are transmitted by its active counterparties. The contribution to the loss of $i$ due to counterparty $j$ is proportional to the level of distress of $j$  (the factor $h_j(t)$) and to the exposures of $i$ towards $j$ relative to its equity (the factor $W_{ij}/E_i(t)$). The matrix with elements $W_{ij}/E_i(t)$ has been named matrix of interbank leverage (\cite{battiston2016leveraging}) because it represents the percentage loss of equity of $i$ that corresponds to a $1\%$ devaluation of its exposure to $j$. 

In the original DebtRank paper, Battiston et al.~\cite{battiston2012debtrank} present a study of US commercial banks, and they show that their algorithm can effectively be used to rank banks in terms of their systemic importance. By showing that relatively small banks can be among the most systemically important, and because of the analogy between DebtRank and centrality measures in networks, they introduced into the debate on systemic risk the idea that some banks might be ``too central to fail''.

\subsection{Extensions}

According to the above formulation, because nodes become inactive after they propagate distress once, losses can flow through a cycle in the network only once. To account for further rounds of propagation, Bardoscia et al.~\cite{bardoscia2015debtrank} derived, from the iteration of the balance sheet identity, the following modified dynamic
\begin{eqnarray}
h_i(t+1) &=& {\rm min} \left\{1, h_i(1) + \sum_{j=1}^N \frac{W_{ij}}{E_i(0)} h_j(t) \right\},\\ 
\end{eqnarray}
where $h_i(1)$ is the initial exogenous shock that affects bank $i$, and it is assumed that $h_i(0)=0$ for all $i = 1,\ldots N$.
This formulation makes it easier to understand the stability of a system with respect to a small perturbation. In particular, if the largest eigenvalue of the matrix of interbank leverages is larger than one, shocks will be amplified by the network and lead to the default of some banks in the system. 

The underlying assumption of DebtRank is that losses propagate from borrowers to lenders linearly: an $x\%$ devaluation of the equity of the borrower leads to an $x\%$ devaluation of the interbank asset of the lender. This assumption can be relaxed by considering dynamics of the form
\be
h_i(t+1) = {\rm min} \left\{1, h_i(1) + \sum_{j=1}^N \frac{W_{ij}}{E_i(0)} f\left( h_j(t) \right)\right\},
\ee
with $f(x)$ a function that maps the interval $[0,1]$ into the positive real semiaxis.
%The DebtRank dynamic has also been extended to the case of a non-linear propagation of distress in 
For instance, Bardoscia et al.~\cite{bardoscia2016distress} considered the following function
\be
f(x) = x e^{-\alpha (x-1)},
\ee
where $\alpha\ge0$. This function represents a one-parameter family of distress propagation rules that interpolates between the threshold model used in \cite{gai2010contagion}, which is recovered in the limit $\alpha\to\infty$, and the linear rule of DebtRank, which corresponds to $\alpha=0$. By performing contagion analyses on a system of European banks, they explored the dependence of the model on the parameter $\alpha$, showing the existence of different regimes for what concerns the amplification of distress.

Bardoscia et al.~\cite{bardoscia2017pathways} also considered the case of a  non-linear propagation of distress, and considered $f(x)$ to be increasing and convex.
They analyzed the largest eigenvalue of the interbank leverage matrix, and they showed the existence of trajectories in the space of networks that can turn a system from stable to unstable through processes that are normally believed to increase the stability of financial markets, namely market integration and diversification. To show this they considered the hypothetical case in which the network of interbank contract is a directed acyclic graph, which is stable. They then considered a situation in which links are randomly added between banks in the system, but in such a way that, every time a link is added, the total amount of lending of each bank is preserved. This implies that banks are on average increasing their diversification. They %\cite{bardoscia2017pathways}
showed that, through this process of increasing diversification, it is possible for an initially stable network to become unstable. They argue that the instability of a network under these type of dynamics is due to the emergence of peculiar cyclical structures in the network of interbank exposures.   

DebtRank is also at the basis of the stress testing framework proposed by Battiston et al.~\cite{battiston2016leveraging}, who propose a framework based on the following steps: 1) application of an exogenous shock to the system and estimation of direct losses; 2) propagation of distress through the DebtRank dynamic Eq.~\eqref{eq:debtRank} and estimation of second-round losses; 3) further (third-round) losses caused by banks liquidating a common asset to target their initial leverage. Through the application of this stress testing framework, Battiston et al.~\cite{battiston2016leveraging} find that the second-round effects (due to DebtRank) and the third-round effects (due to leverage targeting) dominate the first-round losses (direct losses due to the exogenous shock). This finding has potential implications for regulators, as it implies that stress tests that do not account for network effects can significantly underestimate systemic risk. 

In relation to policy making, an interesting work has been presented by Thurner and Polenda~\cite{thurner2013debtrank} and Polenda and Thurner~\cite{poledna2016elimination}, who, using DebtRank as a tool to measure systemic risk, showed how taxation policies that do take into account the impact of interbank contracts on systemic risk can effectively promote systemic stability while not reducing the volume of interbank lending.

\section{Overlapping portfolios and price mediated contagion}\label{sec:overlapping}

\begin{figure}[t]
\begin{center}
           \includegraphics[width=.4\columnwidth]{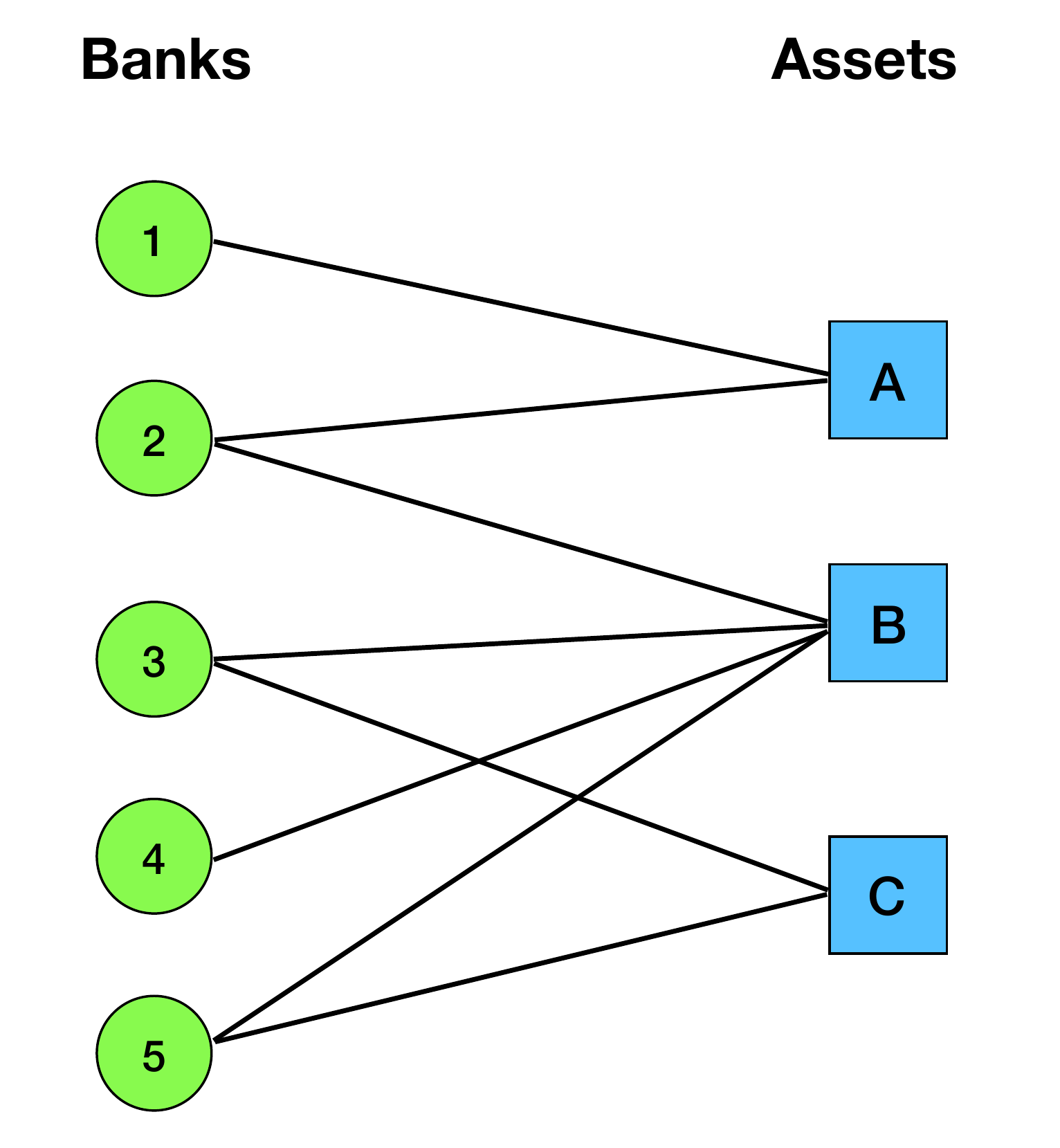}
                \end{center}
    \caption{Pictorial representation of a network of overlapping portfolios. A bank is connected to the assets in its balance sheet. Stress can propagate between banks with common assets. For instance: if bank $\mathtt{1}$ is under distress and liquidates its portfolio, asset ${\mathsf A}$ would be devalued. This would cause a loss to bank $\mathtt{2}$, which might in turn need to liquidate its investment portfolio. This liquidation would cause assets ${\mathsf A}$ to be further devalued and asset ${\mathsf B}$ to be devalued as well, thus causing losses to banks $\mathtt{3}$, $\mathtt{4}$ and $\mathtt{5}$ and the consequent devaluation of asset ${\mathsf C}$.}
 \label{fig:overlappinPortfoliosNetwork}
\end{figure}

Section \ref{sec:interbank_cascades} discussed contagion due to counterparty default risk. Here we discuss a different contagion mechanism, which is associated with the fact that stress propagates between investors that hold common assets. The idea of this contagion mechanism is the following: Consider the simple situation in which two banks $i$ and $j$ invest in a common asset $x$. Suppose now that bank $i$ is under stress, and that in order to reduce its risk exposure it has to liquidate part of its position on asset $x$. Because of market impact, the tendency of prices to react to trading activity, the liquidation procedure causes a devaluation of the asset, whose price will drop. When assets are marked to market, this devaluation causes a loss to bank $j$. We then see that, even in the absence of direct contracts between $i$ and $j$ (such as those associated with interbank loans considered previously), stress can propagate from $i$ to $j$ through the intermediation of the price of the common asset $x$. 
Similarly to the case of counterparty default risk, the question to be asked is therefore ``how does the pattern of overlapping portfolios between banks, which can be modeled as a bipartite network, affect systemic risk``?
A pictorial representation of a simple network of overlapping portfolios is shown in Fig.~\ref{fig:overlappinPortfoliosNetwork}.

The effect of losses due to common asset holdings and fire sales has first be studied by Cifuentes et al.~\cite{cifuentes2005liquidity} in the context of the Eisenberg-Noe model. In their paper, Cifuentes et al.~\cite{cifuentes2005liquidity} consider a system of banks that are interacting through a network of interbank lending relationships and in which all banks are investing in one common external asset. Banks are subject to a capital constraint, and thus need to liquidate part of their investment in the common asset if they face a loss. The authors present numerical simulations on a system of $10$ banks to show the response of the system to the initial default of a bank, and they study the effect of changing the average connectivity of the interbank network. They find that there is a non-monotonic relationship between the number of connections in the network and the number of observed defaults. 
A similar setting is also explored by Gai and Kapadia~\cite{gai2010contagion}, where the interbank lending network is however modeled as a directed \ER network and counterparty default contagion is modeled through the threshold dynamics discussed in section \ref{sec:interbank_cascades}, as well as in Nier et al.~\cite{nier2007network}.
May and Arinaminpathy~\cite{may2010systemic}  build on the model of Nier et al.~\cite{nier2007network}, of which they present a mean-field solution, by considering that banks interact through different asset classes and accounting for contagion between those asset classes. %As for \cite{cifuentes05liquidity,nier2007network,gai2010contagion}, the idea is that a bank in default liquidates its external assets, which results in their depreciation and further losses for other banks holding such assets. The new element they introduce is that in addition to the contagion experienced on the assets that are being liquidated, other banks also suffer a loss associated with  other asset classes

The papers mentioned above considered the effect of fire sales of one or few asset classes, but their focus was the study of the stability of a system as a function of the properties of the network of interbank loans. More recently, the focus shifted towards the study of the network of overlapping portfolios itself and on how its shape affects systemic stability.
The network of overlapping portfolios is usually modeled as a bipartite network, where two types of nodes exist (banks and assets) and a link can only connect a bank to an asset, meaning that the bank is investing in that asset. If we consider a system of $N$ banks and $M$ assets, we can describe the structure of the system in terms of the matrix $Q$, where the element $Q_{ia}$ is the number of shares of asset $a$ held by bank $i$, and we also denote by $p_a$ the price of asset $a$.
Apart from the network, there are two main ingredients that are needed to define a model. The first is the response of the bank to its losses, the second is the response of the asset to its liquidation. If we consider a  dynamic that occurs over discrete time steps $t=1,2,\ldots$, the response of bank $i$ can be defined in terms of the map

\be
Q_{ia}(t) = f_{ia} \left[Q_{ia}(t-1),A_i(t-1),E_i(t-1)\right],
\ee
where we have denoted by $A_i(t)$ the value of the assets of bank $i$ at time $t$ and by $E_i(t)$ its equity,
while the response of asset $a$ can be expressed in terms of the map
\be
p_a(t) = g_a\left[\{Q_{ia}(t)\}\right],
\ee
where we denote by $\{Q_{ia}(t)\}$ the set $\{Q_{1a}(t),Q_{2a}(t),\ldots,Q_{Na}(t)\}$.

In the literature on network models of overlapping portfolios, two choices are common for what concern the response of banks: Either banks are passive until they default, at which point they liquidate their entire portfolio, or they target a certain level of leverage, defined as the ratio between the mark to market value of their assets and their equity:
%\be
%\lambda_i = \frac{A_i}{E_i}.
%\ee
In the following we briefly discuss some of these models. Although with some differences, the models we discuss are all very similar in their ingredients, but the analysis quite different in their focus.

\subsection{Threshold dynamics}

Huang et al.~\cite{huang2013cascading} consider the situation in which a bank is passive until its default, and it liquidates its entire portfolio when it defaults, so that
\be\label{mapfHuang}
f_{ia}\left[Q_{ia}(t-1),A_i(t-1),E_i(t-1)\right]=
\begin{cases}
Q_{ia}(0),~{\rm if}~E_i(t-1)\ge 0\\
0,~{\rm if}~E_i(t-1)< 0
\end{cases}.
\ee
In Huang et al.~\cite{huang2013cascading} asset prices are assumed to respond to liquidation as
\be\label{mapgHuang}
g_a\left[\{Q_{ia}(t)\}\right] = 
p_a(0)\left(1-\alpha\frac{\sum_i\left[Q_{ia}(0)-Q_{ia}(t)\right]}{\sum_i Q_{ia}(0)}\right),
\ee
where $\alpha\ge 0$ is a parameter related to the market impact associated with asset $a$. The above expression means that the value of the asset at time $t$ depends linearly on the fraction of its shares (relative to the total number of shares held in the system) that has been liquidated up to that time.

Huang et al.~\cite{huang2013cascading} perform an empirical analysis concerning the situation of US commercial banks in $2007$. They consider data for $7,\!846$ commercial banks and $13$ asset classes, and they perform stress tests by reducing the value of one asset class from $p_a(0)$ to $(1-\xi) p_a(0)$, with $0\le\xi\le1$. They then compute the number banks that survive the cascading process triggered by the initial shock. They find that abrupt transitions occur in the number of surviving banks as a function of the parameters $\alpha$ and $\xi$ and that devaluation of commercial real estate loans are responsible for the failure of commercial banks during the subprime crisis. Quite interestingly, Huang et al.~\cite{huang2013cascading} also perform an empirical validation of their model by comparing the banks that their model predicts should fail with those banks that actually failed between 2008 and 2011. Their analysis of false positive and true positive rates shows that there is predictive power in the model. 

\begin{figure}[t]
\begin{center}
           \includegraphics[width=.55\columnwidth]{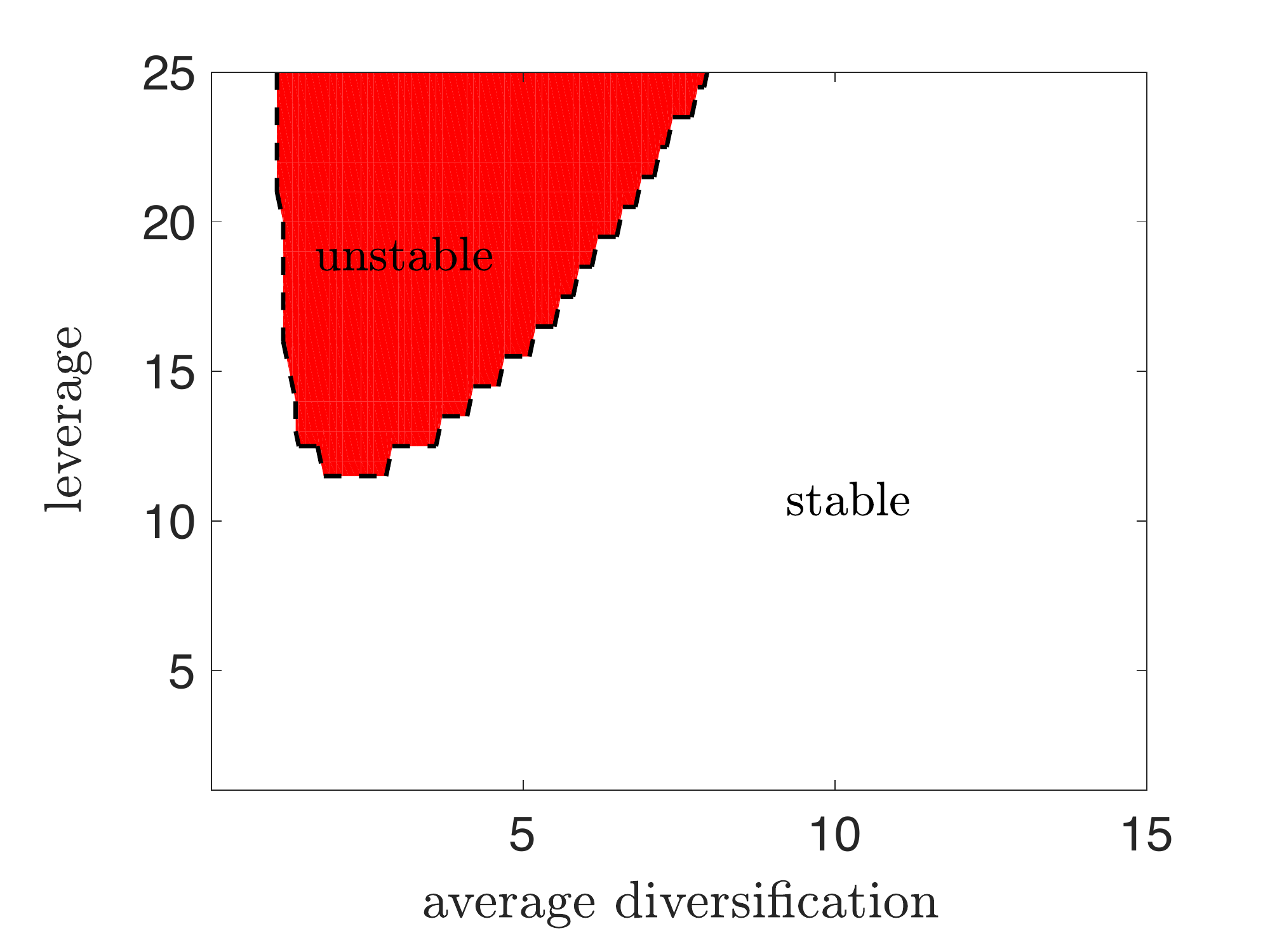}
                \end{center}
    \caption{Region of instability for the cascade model of overlapping portfolios for a bipartite \ER network. Within the red region the system displays global cascades.}
 \label{fig:overlappinPortfoliosModel}
\end{figure}

A similar model of overlapping portfolios is the one of Caccioli et al.~\cite{caccioli2014stability}, who also consider a bipartite network of banks and assets and the map \eqref{mapfHuang} for the update of banks positions on the assets. The rule for the devaluation of assets is however linear in the log returns\footnote{Caccioli et al.~\cite{caccioli2014stability} also test their results with respect to other choices of the market impact function, such as the linear function \eqref{mapgHuang} and a square root market impact law.}, as the one of \cite{cifuentes2005liquidity,gai2010contagion}. This can be written as
\be\label{mapgCaccioli}
g_a\left[\{Q_{ia}(t)\}\right] = 
p_a(0)\left(1-e^{\alpha\frac{\sum_i\left[Q_{ia}(0)-Q_{ia}(t)\right]}{\sum_i Q_{ia}(0)}}\right).
\ee
Caccioli et al.~\cite{caccioli2014stability} study the stability of the system in the limit when the number of banks and assets is large. In particular they identify the conditions under which a small initial perturbation such as the initial bankruptcy of a bank or the devaluation of an asset can lead to a global cascade of bankruptcies. They show that the model can be described in terms of a branching process. In particular, using the approximation that the network is a tree, they define a transfer matrix $\Pi$, whose element $\Pi_{ij}$ represents the probability that the only failure of bank $j$ triggers the failure of bank $i$:
%fails at time step $t$ given the failure of bank $j$ at time $t-1$, 
\be
\Pi_{ij} = {\rm prob}\left[ \sum_{a=1}^M Q_{ia} p_a(0)\left(1 - e^{-\alpha Q_{ja}/\sum_k Q_{ka}} > E_i\right) \right].
\ee
The stability of the system as a function of the model parameters can at this point be assessed by studying the largest eigenvalue of $\Pi$. In the paper, Caccioli et al.~\cite{caccioli2014stability} provide results for networks in the bipartite \ER ensemble, and, similarly to the case of counterparty default risk \cite{gai2010contagion}, they find the existence of a non-monotonic relation between average diversification and probability of observing a global cascades. They also show the existence of a critical value of leverage below which, independently on network connectivity, the system is always stable with respect to the initial shock. This is shown in Fig.~\ref{fig:overlappinPortfoliosModel}, which presents an illustration of the unstable region as a function of the average diversification of banks (i.e. the average degree of banks in the network of overlapping portfolios) and the leverage of banks, which is defined as the market value of the bank's investment portfolio divided by its equity. The figure refers to the same setting considered in \cite{caccioli2014stability}, with a bipartite \ER networks, and under the assumption that all banks have the same leverage.   
The effect of heterogeneous degree distributions on this model is studied in Banwo et al.~\cite{banwo2016effect} by means of numerical simulations.

\subsection{Leverage targeting}

%\tk{I replaced some pointers \cite{caccioli2014stability} with ``they" to avoid repetitive usage.}

Caccioli et al.~\cite{caccioli2014stability} also consider the effect of relaxing the assumption that banks are passive investors, and look at what happens if banks decide to target their initial leverage over the dynamics. If a bank suffers a loss, its leverage will go up, therefore a rebalancing is needed to reduce risk. %\cite{caccioli2014stability}
They consider the situation in which banks react to losses at time $t$ by liquidating a fraction
\be
\Delta A_i(t) = \gamma A_i(t)\left(1-\frac{\lambda_iE_i(t)}{A_i(t)}\right)
\ee
of their investment. In the above formula, $\lambda_i$ is the target leverage of bank $i$, while the parameter $\gamma\in[0,1]$ determines how quickly the bank tries to reach its target. $\Delta A_i(t)$ is the total value of the assets liquidated by the bank at time $t$. To know how many shares of a given asset $a$ are sold, one needs to divide by the number  $k_i$ of different assets in the bank's portfolio (it is assumed here that banks liquidate the same fraction of each asset), and by the current price of asset $a$.
This leads to the following response function
\be\label{mapfCaccioli}
f_{ia}\left[Q_{ia}(t-1),A_i(t-1),E_i(t-1)\right]=
\begin{cases}
Q_{ia}(t-1)-\gamma \frac{A_i(t-1)}{k_i p_a(t-1)}\left(1-\frac{\lambda_i E_i(t-1)}{A_i(t-1)}\right),~{\rm if}~E_i(t-1)\ge 0\\
0,~{\rm if}~E_i(t-1)< 0
\end{cases}.
\ee
%\cite{caccioli2014stability} 
They find that the attempt of banks to reduce their individual risk through preemptive liquidation ends up significantly widening the region of parameter space where global cascades can be observed.%\tk{I think the cascade region is the parameter space within which a global cascade CAN occur.}

%Leverage targeting is a common assumption of contagion models because \cite{adrian2009liquidity} showed empirical evidence that commercial banks do indeed target a constant leverage, and leverage targeting is equivalent for an investor to maximizing its expected return on equity given a VaR constraint \cite{shin2010risk}. 
%The fact that individual risk management based on VaR can lead to systemic consequences is well demonstrated in \cite{corsi2016when}. They consider a system of N banks invests in $m$ randomly chosen assets out of a universe of $M$ possible assets, where $m$ is however computed as the optimal value of diversification that corresponds to banks maximizing their profit conditional on a VaR constraint. The value of the assets is then the sum of a common factor and idiosyncratic random variables, plus a linear market impact term that depends on banks trading. As in the other models we discuss here, the relative composition of the portfolio does not change over time, but at each time banks change the volume of their investment on the portfolio in order to maintain their target leverage. Quite interestingly, in their dynamical model they show that, upon reducing the diversification cost, the system goes from a stable to an unstable regime.

A stress test framework based on shock propagation due to leverage targeting is the one proposed by Greenwood et al.~\cite{greenwood2015vulnerable}. In this case banks estimated their losses at time $t$ and then reduce the size of their investment in order to restore their initial leverage at time $t+1$, which corresponds to the map given in equation \eqref{mapfCaccioli} with $\gamma=1$. Deleveraging causes prices to be devalued, and consequently mark-to-market losses for banks, that will further need to deleverage, and so on. The market impact function considered by \cite{greenwood2015vulnerable} is a linear function like the one in equation \eqref{mapfHuang}. They try to disentangle the effect of different factors on the losses produced by fire sale contagion, showing that the contribution of a bank to aggregate deleveraging is the higher if a bank is more connected, bigger, more leveraged and more exposed to the initial shock. %\cite{greenwood2015vulnerable}
They also introduce the concept of indirect vulnerability of a bank with respect to a given asset, as measured by the loss of the bank's equity due to the deleveraging associated with the asset. This has to be contrasted with the notion of direct vulnerability, which is the direct exposure of the bank towards the asset. They test the model on the largest $90$ banks in the European Union for the period 2009-2011. Through a regression analysis on those banks that are publicly traded, they find that direct and indirect vulnerabilities have the same explanatory power on banks' returns.

The model of  Greenwood et al.~\cite{greenwood2015vulnerable} is at the basis of the framework developed in Duarte and Eisenbach~\cite{duarte2015fire} to measure aggregate vulnerability and systemic importance of banks.   Greenwood et al.~\cite{greenwood2015vulnerable} also show that their measure of aggregate vulnerability due to spillover effects grows well before the crisis of 2008, and they are able to disentangle the contribution to aggregate vulnerability due to the increase of leverage, system size and concentration of investments in illiquid assets.

Cont and Schaanning~\cite{cont2017fire} consider a dynamic which is in between that of a passive investor and a leverage targeting one. In their model, they account for the fact that there is usually a buffer between the leverage of a bank and the maximum leverage allowed by regulation. This is done so that they are not forced to liquidate their position because of a relatively small loss. %\cite{cont2017fire}
They consider then the bank as a passive investor until its loss makes the bank break its leverage constraint. When this happens, the bank develerages in order to reach its target, which is however below the maximum allowed by the regulatory regime (so that a small buffer is restored). %\cite{cont2017fire} 
They also distinguish between marketable securities, that can be liquidated and are subject to market impact, and illiquid assets that are not marketable and therefore cannot be liquidated. Deleveraging only involves marketable securities.  

Cont and Schaanning~\cite{cont2017fire} consider data collected by the European Banking Authority on $51$ European banks, and they compare the outcome of stress tests performing with their model vs. target leveraging. In terms of price changes, they consider both a linear and square root market impact. They find that there are significant differences in the losses estimated with the two models. Quite interestingly, they also introduce a matrix of overlaps between the portfolios of different banks, where each asset is weighted by its liquidity, and they show that, although many financial institutions have zero overlap between their portfolios, they are all connected by second order overlaps. This means that stress tests that do not account for second round losses can significantly underestimate systemic risk.

%They also introduce the notion of  indirect vulnerability of a bank $i$ towards another. bank $j$ by computing the loss induced to $i$ should $j$ deleverage, finding that indirect vulnerability is high if two banks own common illiquid assets.
%They frame their model in terms of two period model in which banks that are subject to a loss at time $t=1$ deleverage at time $t=2$ in order to restore their initial leverage. This leads to mark-to-market losses due to market impact. The model can then be iterated over time 

%Leverage targeting is a common assumption of contagion models because \cite{adrian2009liquidity} showed empirical evidence that commercial banks do indeed target a constant leverage, and leverage targeting is equivalent for an investor to maximizing its expected return on equity given a VaR constraint \cite{shin2010risk}. 
Although there may be randomness in the shocks that hit the system, or in the construction of the network of overlapping portfolios, the dynamics described above are all deterministic.
A stochastic dynamic is for instance the one considered in Corsi et al.~\cite{corsi2016when}. They consider a system of $N$ banks investing in $m$ randomly chosen assets out of a universe of $M$ possible assets, where $m$ is however computed as the optimal value of diversification that corresponds to banks maximizing their profit conditional on a VaR constraint, which is equivalent to leverage targeting \cite{shin2010risk}.
The value of the assets is then the sum of a linear market impact term that depends on banks trading plus a stochastic component, which is in turn the sum of a common factor and an idiosyncratic component.
As for the other models we discuss here, the relative composition of the portfolio does not change over time, but at each time banks change the volume of their investment on the portfolio in order to maintain their target leverage. Corsi et al.~\cite{corsi2016when} show that, upon increasing diversification, the system goes from a stable regime where time series of asset returns are stationary to an unstable regime where they are characterized by bubbles and bursts.\\

%\subsection{Empirical analysis of overlapping portfolios networks}
%braveman empirical\\

%minca empirical\\
 
%Lillo reconstruction\\

%cimini squartini reconstruction\\

\section{Empirical structure of interbank networks}\label{sec:structure}

%Networks of financial relations between institutions emerge from a wide
%variety of markets and display multiple features. 
%In the theory of complex
%systems, the evolution of a system is dependent on a system's emerging structure. These structures derive from a limited set of first principles of the nature of interaction between the agents in the market.

There are many works that aim to characterize the structure of real-world financial networks. 
Measuring and analyzing the structure of financial networks has a two-fold objective: On one hand,  knowledge of the structure of financial networks  gives insights on how local risks would spread over the entire network through financial linkages.
This class of studies aims to measure the systemic risks of particular network topologies that could emerge and remain unchanged for a certain period of time (i.e., static structure). On the other hand, since the topology changes over time, knowledge about the dynamical transition patterns could allow us to predict how systemic risk would evolve over time. 
 In this section, we provide a brief review of these two lines of research that study the static and dynamic structures of interbank networks.

 %Before proceeding, we remark about the availability of interbank data. Since edges of interbank networks are defined by financial interactions between banks (and other financial institutions, more generally), we need to have the data on bilateral transactions to study the empirical network structure. Unfortunately, such data are usually not made public due to confidential reasons, and therefore a large fraction of existing works are done by central bankers and the staffs of financial authorities. One exception is the data for the Italian interbank market, in which banks trade through an online platform called e-MID.\footnote{The e-MID data are commercially available from e-MID S.p.A based in Milan~\cite{emidHP}.}

\subsection{Static structure}

\subsubsection{Interbank networks in different countries}

Over the past decade, the topology of interbank networks has been examined in many countries. These studies include Boss et al.~\cite{Boss2004} for Austria, Upper and Worms~\cite{Upper2004EER} for Germany, Degryse et al.~\cite{Degryse2007IJCB} for Belgium, Van Lelyveld and Liedorp~\cite{Lelyveld2006IJCB} for Netherland, Iori et al.~\cite{Iori2008JEDC} and Bargigli et al.~\cite{bargigli2015multiplex} for Italy, Wells~\cite{wells2004financial} and Langfield et al.~\cite{langfield2014mapping} for the UK, Furfine~\cite{furfine2003interbank} for the US, Cont et al.~\cite{Cont2013} for Brazil, Martinez-Jaramillo et al.~\cite{martinez2014empirical} for Mexico, and Imakubo and Soejima~\cite{Imakubo2010BOJ} for Japan.

 While in some countries bilateral transactions data are available,\footnote{One example is the data for the Italian interbank market, in which banks trade through an online platform called e-MID. The e-MID data are commercially available from e-MID S.p.A based in Milan (\url{http://www.e-mid.it/}).} in many countries the aggregate balance-sheet data (e.g., total amount of loans) are the only source of information for bilateral trades. In such cases, one needs to estimate the interbank network structure using a suitable estimation method. A widely used one is the maximum entropy (ME) method. The ME method estimates the network structure by maximizing the entropy of interbank linkages, which implies that the total interbank lending is distributed to all the possible borrowers as evenly as possible. A disadvantage of the ME method is that it is likely that the estimated network is much denser than the actual one. Mistrulli~\cite{Mistrulli2011JBF} argues that the ME method may over- or underestimate the risk of default contagion. To overcome the problem, recently more sophisticated methods are also proposed~\cite{mastrandrea2014enhanced}. Anand et al.~\cite{ANAND2017} compare the accuracy of several existing estimation methods by applying them to various empirical networks.
 
  A more direct way to extract information of bilateral transactions is to use interbank payment data. Since payment flows contain information on interbank settlements and transfers, one could filter out the information of bilateral interbank loans. This approach is taken by, among others, Furfine~\cite{furfine2003interbank}, Demiralp et al.~\cite{demiralp2006overnight}, and Imakubo and Soejima~\cite{Imakubo2010BOJ}.

 \subsubsection{Core-periphery structure}
 
 It has been argued that the structure of interbank networks at certain point in time is best described as a core-periphery structure~\cite{borgatti2000coreperiphery}. A core-periphery structure is formed by two groups: \emph{core} and \emph{periphery}. The core and peripheral nodes are distinguished as follows; the core forms a subgraph of the entire network in which nodes are connected densely to each other. Peripheral nodes are connected to the core nodes but not to other peripheral nodes. This is expressed by a block adjacency matrix as
 \begin{align}
     A = \left[ \begin{matrix} \mathbf{CC} \;& \mathbf{CP} \\ \mathbf{PC}& \mathbf{PP}\end{matrix}\right] 
 \approx \left[ \begin{matrix} \mathbf{1} \;& \mathbf{CP} \\ \mathbf{PC}& \mathbf{0}\end{matrix}\right],
 \label{eq:core_peri}
 \end{align}
 where $\mathbf{CC}$ denotes the submatrix representing the connectivity among core nodes, and $\mathbf{PC}$ represents the connectivity between the core and peripheral nodes. $\mathbf{PC}$ is identical to $\mathbf{CP}$ since we consider undirected graphs. In general, detecting core-periphery structure in a directed graph is a challenging problem, and most of the existing methods for core-periphery detection are developed for undirected graphs. In the pure core-periphery structure, we should have $\mathbf{CC}=\mathbf{1}$ (i.e., a complete graph) and $\mathbf{PP}=\mathbf{0}$ (i.e., there is no link between peripheral nodes).
 %\footnote{Lux~\cite{lux2015emergence} proposed a model that explains the emergence of core-periphery structure in interbank markets.}
  Classifying all the nodes into core and periphery is a nontrivial task, and a popular way to do this is to find core nodes so as to minimize the difference between the empirical adjacency matrix and the ideal core-periphery block matrix~\eqref{eq:core_peri}~\cite{borgatti2000coreperiphery,craig2014interbank,Fricke2015CompEcon}. 
  
  %This method is based on the idea that the classification of core and peripheral nodes should be most plausible when the distance between the empirical adjacency matrix and the ideal block matrix is minimized.

  \subsubsection{Empirical networks: core-periphery vs. bipartite structure}
  
  We summarize the works estimating the core-periphery structure in interbank networks in Table~\ref{tab:structure}. These empirical studies differ in terms of the data and their time scales. For example, Fricke and Lux~\cite{Fricke2015CompEcon} and Barucca and Lillo~\cite{Barucca2016Soliton,Barucca2017Comp} used the data for interbank transactions, but the former studied quarterly aggregate networks and the latter analyzed daily networks. Imakubo and Soejima~\cite{Imakubo2010BOJ} studied the transactions data filtered out of interbank payment data, and other studies are based on the regulatory data reported by financial institutions to the financial authorities.

  While many empirical works on core-periphery structure are essentially based on the standard detection method described above, a more flexible and widely used approach to detecting block structure, called the stochastic block model (SBM)~\cite{peixoto2017bayesian}, has also been used. The SBM is a probabilistic model of random graphs with a flexible block structure; nodes are assigned to different blocks and each pair of nodes is linked with a probability depending on the nodes' blocks. This model can generate arbitrary block structures, such as core-periphery, modular, and bipartite structures.
  In fact, Barucca and Lillo~\cite{Barucca2016Soliton,Barucca2017Comp} employ this approach and find that the two-block structure that best represents the e-MID overnight money market is bipartite (i.e. borrowers and lenders) at the daily resolution.

\begin{figure}[t]
      \centering
      \includegraphics[width=.75\columnwidth]{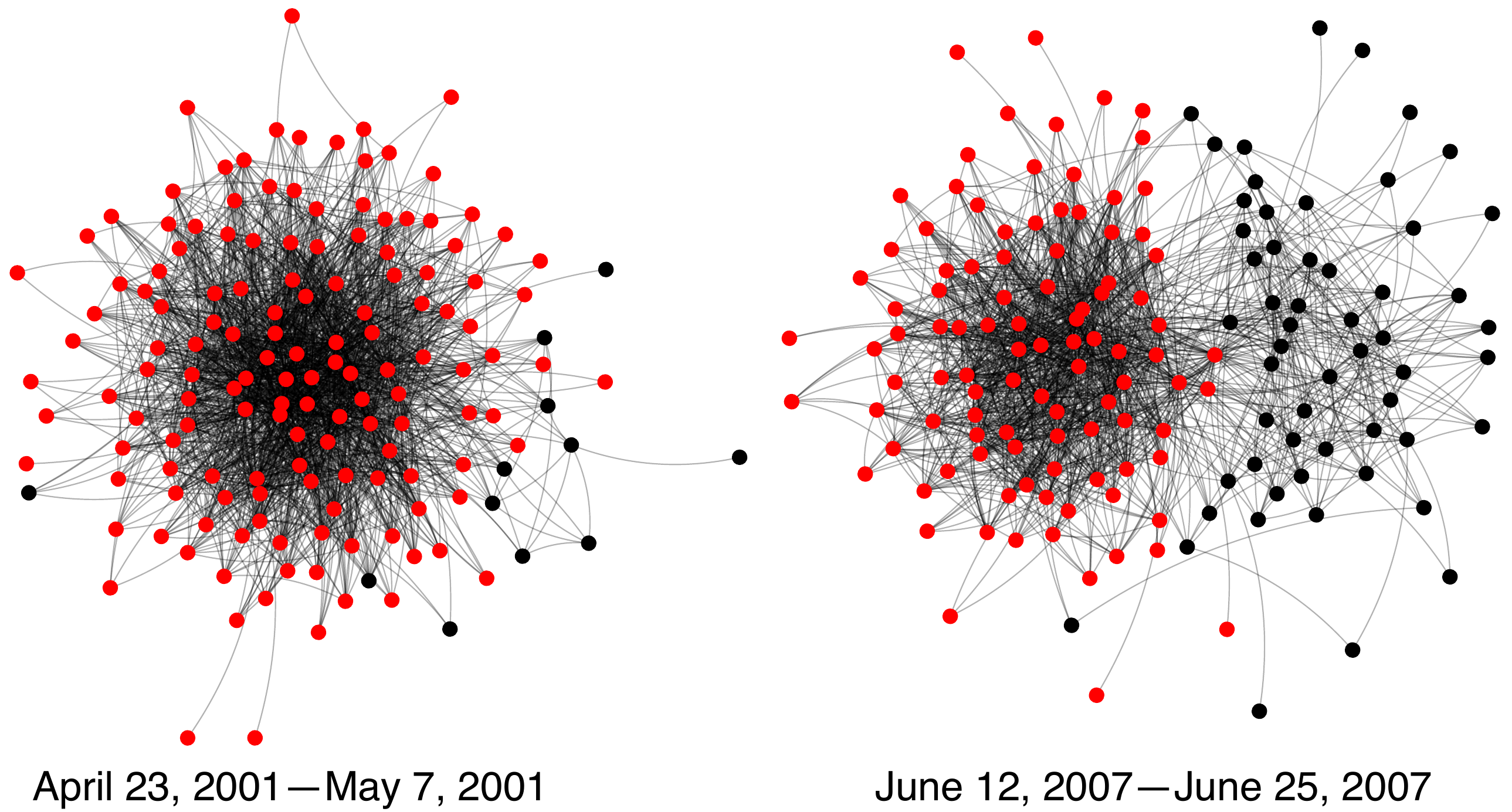}
      \caption{Visualization of Italian interbank networks, e-MID. The networks are aggregated over 10 business days. Red and black circle denote Italian and other foreign banks, respectively. Visualization is done by python-igraph with the Kamada-Kawai algorithm~\cite{Kamada1989algorithm}.}
      \label{fig:visualization}
  \end{figure}

 Even in the case of aggregate networks, the plausibility of the stylized core-periphery structure  as a characteristic of interbank networks can be controversial. As is evident from Eq.~\eqref{eq:core_peri}, the previous works implicitly assumed that an empirical network consists of a single core block and a single peripheral block. This suggests that even if there were no such a standard core-periphery structure in the empirical network, the estimation method classifies each node as either a core node or a peripheral node. In fact, recent studies argue that the seemingly core-periphery structure might just come from a heterogeneous degree distribution or consist of two cores~\cite{rombach2014core,kojaku2017finding,kojaku2017core}.  
  As an example, visualization of Italian interbank networks aggregated over 10 business days is presented in Fig.~\ref{fig:visualization}. It appears that there are densely connected core nodes at the center of the network while there are also (seemingly) peripheral nodes that are not linked to each other yet connected to the core. However, in 2007, Italian banks and other foreign banks seem to form two core-like groups, which would make it difficult to extract a stylized pure core-periphery structure. It might be reasonable to infer that there are two core-periphery structures in the network~\cite{kojaku2017finding}.

\begin{table}[]
    \centering
    \begin{tabular}{lcccc}
    \hline 
    \hline
        Reference & Data type & Country & Resolution & Year \\
        \hline 
        Fricke and Lux~\cite{Fricke2015CompEcon} & transactions data & Italy & quarterly &2015 \\
        Craig and von\ Peter~\cite{craig2014interbank}  & regulatory data & Germany & quarterly & 2014 \\
        Veld and van Lelyveld~\cite{van2014finding} & regulatory data &Netherlands & quarterly &2014\\
        Langfield et al.~\cite{langfield2014mapping} & regulatory data &UK & annual & 2014\\
        Imakubo and Soejima~\cite{Imakubo2010BOJ} & settlement data & Japan & monthly & 2010 \\
        Barucca and Lillo~\cite{Barucca2016Soliton,Barucca2017Comp} &transactions data & Italy & daily &2016\\
        \hline
    \end{tabular}
    \caption{Empirical studies on the core-periphery structure in interbank networks.}
    \label{tab:structure}
\end{table}

% \tk{Papers suggesting core-periphery: \\
% Veld and van Lelyveld (J.Bank.Fin.2014, Dutch data),\\
% Craig and von Peter (2014, J.Fin.Intermediation, German data),\\ 
% Lux (JEDC, 2015, modelling the emergence of core-periphery,)\\
% Langfield et al(2014,JBF, UK data, interbank exposure and interbank funding nets)\\
% Sang Hoon Lee, Mihai Cucuringu, and Mason A. Porter (PRE 2014, 032810. Data from the European Banking Authority report)
% }

\subsection{Interbank network dynamics at the daily scale}

\subsubsection{Why daily scale?}

 Although the majority of empirical works are based on static and aggregate networks, the granularity of the data can be crucial for recognizing the heterogeneous behavior of financial institutions and for capturing the functioning of a market at its inherent time-scales. In interbank markets, most of the bilateral transactions are overnight, meaning that the relationship between two banks as a lender and a borrower lasts only for one day or shorter, depending on the time when the loan contract is made. In the Italian interbank market, for example, more than 86\% of transactions are overnight lending in the period between 2000--2015~\cite{Kobayashi2017arxiv}.
 
 If the issue of interest is to understand the interconnectedness of financial risks, then an aggregate network of overnight lending relationships contains irrelevant information because different edges formed in different days in fact do not exist at the same time. Many researchers studied aggregate networks not necessarily because it conveys information about the interconnected risk structure, but rather because it would reveal meaningful information about the structure of long-term relationships among banks. Another possible reason for why aggregate networks attract attention is that daily networks can be much sparser and noisier than aggregate networks, and they appear to change their structure from day to day in a purely random manner~\cite{finger2013network,Musmeci2013JSP}.

 \subsubsection{Daily network dynamics}
 
  Barucca and Lillo~\cite{Barucca2016Soliton,Barucca2017Comp} show that at the daily resolution, the structure of interbank networks in the e-MID interbank market is not characterized by a core-periphery structure, but rather characterized by a bipartite structure or more general community structures. Lowering the time resolution (such as weeks or months) tends to increase the likelihood that a core-periphery structure is detected, yet a non-negligible fraction of such networks are still best characterized by a bipartite structure.
  Kobayashi and Takaguchi~\cite{Kobayashi2017arxiv} reinforce their result by showing that the \emph{bipartivity} of daily interbank networks have been increasing over the past decade.\footnote{Bipartivity is a measure of bipartite structure proposed by Estrada and Rodr\'{\i}guez-Vel\'azquez~\cite{Estrada2005PRE}, which takes 0.5 if the network is complete and one if purely bipartite.}

  \begin{figure}[t]
      \centering
      \includegraphics[width=.97\columnwidth]{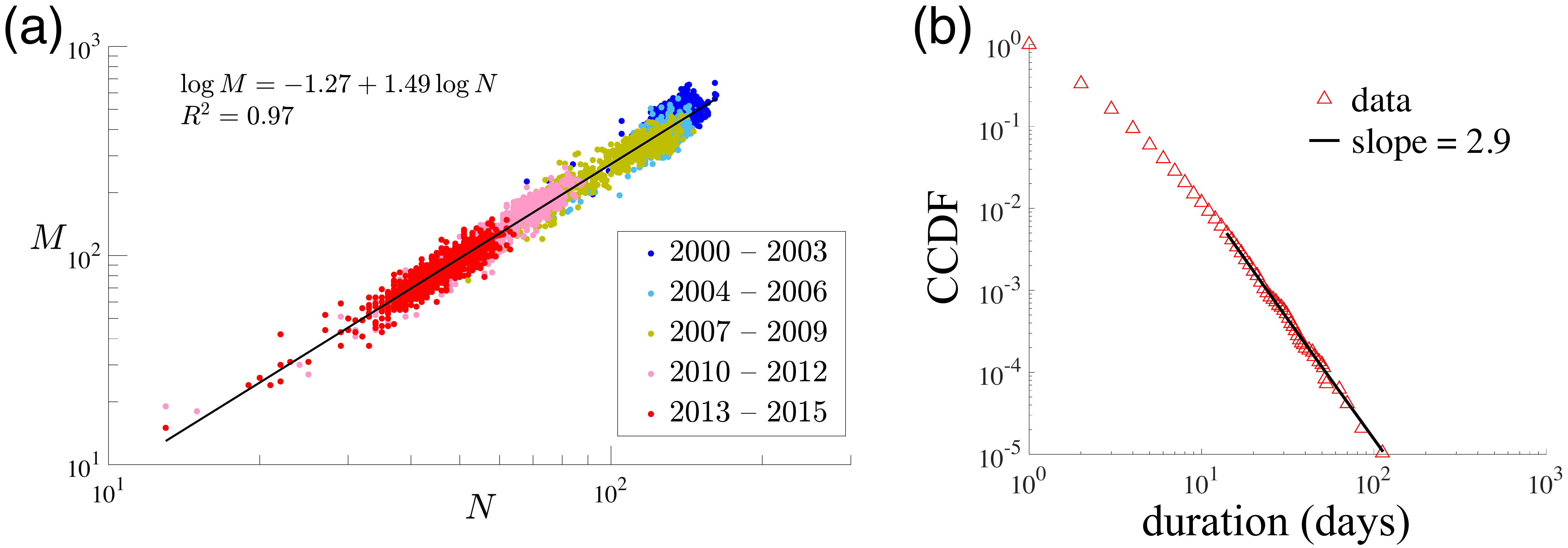}
      \caption{Scaling laws in the Italian overnight interbank networks. (a) Superlinear relationship between the numbers of nodes ($N$) and edges ($M$) between September 4, 2000 and December 31, 2015 (3922 business days). Each dot corresponds to a day. (b) Complementary cumulative distribution function (CCDF) of transaction duration (in terms of the number of business days) of bank pairs between 2010--2015.}
      \label{fig:scaling}
  \end{figure}

  If we regard interbank markets as dynamic systems in which the structure of bilateral exposures changes every day, an interesting question to ask is whether the daily dynamics are just random or there are robust and time-invariant properties. Kobayashi and Takaguchi~\cite{Kobayashi2017arxiv} find that the daily market activity represented by combination $(N,M)$, where $N$ and $M$ are the numbers of active banks and edges, respectively, is strictly ruled by a superlinear relationship $N\propto M^{1.5}$, or $\langle N\rangle \propto \sqrt{M}$, independently of the structure and the size of daily networks (Fig.~\ref{fig:scaling}a). They also find several daily dynamical patterns in the e-MID market, such as the power-law distribution of transaction duration (Fig.~\ref{fig:scaling}b) and a tent-shaped distribution of weight growth. Interestingly, these properties are ubiquitous in social networks formed via human interactions such as phone calls and face-to-face interactions~\cite{Cattuto2010PLOS,Starnini2013PRL,Schlapfer2014}.

% \subsection{Heterogeneity and concentration}
% Cont-Moussa
% Iori-Caldarelli 

% \subsection{Community organization}
% Core-periphery Fricke-Lux vs bipartite structure Kobayashi, Lillo-Barucca

\section{Discussion}\label{sec:discussion}

 In this article, we reviewed recent works studying financial systemic risk based on network approaches. 
 While we tried to cover as many research topics as possible, these are obviously not exhaustive. In particular, two important research areas that have not been discussed in the current article are the prediction and the control of systemic risk. 
 
 As in other fields of sciences, there are generally three steps for the study of systemic risk to mature as a scientific research field: The first step is to understand and model the mechanisms behind the real-world phenomena. The second is to forecast the future state of the system. The third and last step is to control the system to avoid the occurrence of undesired phenomena. Most of the studies discussed in the current article are still at the first stage. Researchers from various fields have just started working together since the late 2000's to develop  models, such as the models of interbank default cascades through bilateral exposures and overlapping portfolios, that can be used to describe the real-world phenomena analytically.  
 
 In recent years, however, a growing number of researchers are tackling the controllability of systemic risk by simulating possible policy tools that could be taken by the financial regulators.
 We still need further studies at every step to deepen our understanding of
the complexity of financial networks and reduce systemic risk. We hope that this review article will encourage researchers from various areas of sciences to join in this challenging research field.

\section*{Acknowledgements}
F.C. acknowledges support
of the Economic and Social Research Council (ESRC) in funding the Systemic Risk
Centre (ES/K002309/1). P.B. acknowledges financial support from FET Project DOLFINS nr.\ 640772.
T.K. acknowledges financial support from the Japan Society for the Promotion of Science Grants no.\ 15H05729 and 16K03551.

%\begin{acknowledgements}\tk{(non-financial support)}
% P.B. acknowledges financial support from DOLFINS (number to be specified).
% T.K. acknowledges financial support from the Japan Society for the Promotion of Science Grants no. 15H05729 and 16K03551.
%\end{acknowledgements}

%\bibliographystyle{unsrt}
%\bibliographystyle{spmpsci}
%\bibliographystyle{plain}
%\bibliographystyle{spbasic}

%\bibliographystyle{spbasic_unsort}
%\bibliography{interbank_survey}   % name your BibTeX data base

\begin{thebibliography}{98}
\providecommand{\natexlab}[1]{#1}
\providecommand{\url}[1]{{#1}}
\providecommand{\urlprefix}{URL }
\expandafter\ifx\csname urlstyle\endcsname\relax
  \providecommand{\doi}[1]{DOI~\discretionary{}{}{}#1}\else
  \providecommand{\doi}{DOI~\discretionary{}{}{}\begingroup
  \urlstyle{rm}\Url}\fi
\providecommand{\eprint}[2][]{\url{#2}}

\bibitem[{May et~al(2008)May, Levin, and Sugihara}]{May2008Nature}
May R.~M., Levin S.~A., Sugihara G. (2008) Complex systems: Ecology for
  bankers. Nature 451(7181),~893--895.

\bibitem[{Schweitzer et~al(2009)Schweitzer, Fagiolo, Sornette, Vega-Redondo,
  Vespignani, and White}]{Schweitzer2009Science}
Schweitzer F., Fagiolo G., Sornette D., Vega-Redondo F., Vespignani A., White
  D.~R. (2009) Economic networks: The new challenges. Science
  325(5939),~422--425.

\bibitem[{Battiston et~al(2016)Battiston, Farmer, Flache, Garlaschelli,
  Haldane, Heesterbeek, Hommes, Jaeger, May, and
  Scheffer}]{Battiston2016Science}
Battiston S., Farmer J.~D., Flache A., Garlaschelli D., Haldane A.~G.,
  Heesterbeek H., Hommes C., Jaeger C., May R., Scheffer M. (2016) Complexity
  theory and financial regulation. Science 351(6275),~818--819.

\bibitem[{Haldane and May(2011)}]{haldane2011systemic}
Haldane A.~G., May R.~M. (2011) Systemic risk in banking ecosystems. Nature
  469(7330),~351--355.

\bibitem[{Fouque and Langsam(2013)}]{Fouque2013handbook}
Fouque J.-P., Langsam J.~A. (2013) \emph{Handbook on Systemic Risk}. Cambridge
  University Press.

\bibitem[{Eisenberg and Noe(2001)}]{eisenberg2001systemic}
Eisenberg L., Noe T.~H. (2001) Systemic risk in financial systems. Management
  Science 47(2),~236--249.

\bibitem[{Gai and Kapadia(2010)}]{gai2010contagion}
Gai P., Kapadia S. (2010) Contagion in financial networks. Proceedings of the Royal Society A 466(2120), 2401--2423.

\bibitem[{Watts(2002)}]{Watts2002}
Watts D.~J. (2002) A simple model of global cascades on random networks. Proceedings of the National Academy of Sciences USA 99(9),~5766--5771.

\bibitem[{Battiston et~al(2012)Battiston, Puliga, Kaushik, Tasca, and
  Caldarelli}]{battiston2012debtrank}
Battiston S., Puliga M., Kaushik R., Tasca P., Caldarelli G. (2012) Debtrank: Too central to fail? financial networks, the fed and systemic risk. Scientific Reports 2, 541.

\bibitem[{Battiston et~al(2016)Battiston, Caldarelli, May, Roukny, and
  Stiglitz}]{Battiston2016PNAS}
Battiston S., Caldarelli G., May R.~M., Roukny T., Stiglitz J.~E. (2016) The price of complexity in financial networks. Proceedings of the National Academy of Sciences USA 113(36),~10031--10036.

\bibitem[{Tarski(1955)}]{tarski1955lattice}
Tarski A. (1955) A lattice-theoretical fixpoint theorem and its applications. Pacific Journal of Mathematics 5(2),~285--309.

\bibitem[{Rogers and Veraart(2013)}]{Rogers2013}
Rogers L.~C., Veraart L.~A. (2013) Failure and rescue in an interbank network. Management Science 59(4),~882--898.

\bibitem[{Visentin et~al(2016)Visentin, Battiston, and
  D'Errico}]{visentin2016rethinking}
Visentin G., Battiston S., D'Errico M. (2016) Rethinking financial contagion. arXiv:1608.07831.

\bibitem[{Elsinger et~al(2006)Elsinger, Lehar, and Summer}]{Elsinger2006}
Elsinger H., Lehar A., Summer M. (2006) Risk assessment for banking systems.
  Management Science 52(9),~1301--1314.

\bibitem[{Barucca et~al(2016)Barucca, Bardoscia, Caccioli, D'Errico, Visentin,
  Battiston, and Caldarelli}]{barucca2016network}
Barucca P., Bardoscia M., Caccioli F., D'Errico M., Visentin G., Battiston S., Caldarelli G. (2016) Network valuation in financial systems. arXiv:1606.05164.

\bibitem[{Gleeson and Cahalane(2007)}]{Gleeson2007}
Gleeson J.~P., Cahalane D.~J. (2007) Seed size strongly affects cascades on random networks. Physical Review E 75(5),~056103.

\bibitem[{Lee et~al(2014)Lee, Brummitt, and Goh}]{Lee2014PRE}
Lee K.-M., Brummitt C.~D., Goh K.-I. (2014) {Threshold cascades with response heterogeneity in multiplex networks}. Physical Review E 90(6) 062816.

\bibitem[{Kobayashi(2015)}]{KobayashiPRE2015}
Kobayashi T. (2015) Trend-driven information cascades on random networks. Physical Review E 92(6),~062823.

\bibitem[{Newman(2010)}]{Newman2010book}
Newman M. E.~J. (2010) \emph{Networks: An Introduction}. Oxford University Press.

\bibitem[{Callaway et~al(2000)Callaway, Newman, Strogatz, and
  Watts}]{Callaway2000PhysRevLett}
Callaway D.~S., Newman M.~E., Strogatz S.~H., Watts D.~J. (2000) Network robustness and fragility: Percolation on random graphs. Physical Review Letters 85(25),~5468.

\bibitem[{Kobayashi and Takaguchi(2017)}]{Kobayashi2017arxiv}
Kobayashi T., Takaguchi T. (2017) Social dynamics of financial networks. arXiv:1703.10832.

\bibitem[{Bogu\~n\'a and Serrano(2005)}]{Boguna2005PRE}
Bogu\~n\'a M., Serrano M.~A. (2005) Generalized percolation in random directed networks. Physical Review E 72(1),~016106.

\bibitem[{Payne et~al(2011)Payne, Harris, and Dodds}]{Payne2011PRE}
Payne J., Harris K., Dodds P. (2011) {Exact solutions for social and biological contagion models on mixed directed and undirected, degree-correlated random networks}. Physical Review E 84(1),~016110.

\bibitem[{Hurd and Gleeson(2013)}]{Hurd2013JCN}
Hurd T.~R., Gleeson J.~P. (2013) On Watts' cascade model with random link weights. Journal of Complex Networks 1(1),~25--43.

\bibitem[{Hurd(2016)}]{Hurd2016book}
Hurd T.~R. (2016) \emph{Contagion!: Systemic Risk in Financial Networks}. Springer.

\bibitem[{Unicomb et~al(2017)Unicomb, I{\~n}iguez, and
  Karsai}]{Unicomb2017threshold}
Unicomb S., I{\~n}iguez G., Karsai M. (2017) Threshold driven contagion on weighted networks. arXiv:1707.02185.

\bibitem[{Gleeson(2011)}]{GleesonPRL2011}
Gleeson J.~P. (2011) High-accuracy approximation of binary-state dynamics on networks. Physical Review Letters 107(6),~068701.

\bibitem[{Gleeson(2013)}]{GleesonPRX2013}
Gleeson J.~P. (2013) Binary-state dynamics on complex networks: Pair
  approximation and beyond. Physical Review X 3(2),~021004.

\bibitem[{Erd\H{o}s and R\'enyi(1959)}]{Erdos1959PublMath}
Erd\H{o}s P., R\'enyi A. (1959) On random graphs I. Publicationes Mathematicae 6,~290--297.

\bibitem[{Boss et~al(2004)Boss, Elsinger, Summer, and Thurner}]{Boss2004}
Boss M., Elsinger H., Summer M., Thurner S. (2004) Network topology of the interbank market. Quantitative Finance 4(6),~677--684.

\bibitem[{Iori et~al(2008)Iori, De~Masi, Precup, Gabbi, and
  Caldarelli}]{Iori2008JEDC}
Iori G., De~Masi G., Precup O.~V., Gabbi G., Caldarelli G. (2008) A network analysis of the Italian overnight money market. Journal of Economic Dynamics and Control 32(1),~259--278.

\bibitem[{Cont et~al(2013)Cont, Moussa, and Santos}]{Cont2013}
Cont R., Moussa A., Santos E.~B. (2013) Network structure and systemic risk in banking systems. In: Fouque J.-P., Langsam J.~A. (eds) \emph{Handbook on Systemic Risk}, Cambridge University Press, New York.

\bibitem[{Melnik et~al(2011)Melnik, Hackett, Porter, Mucha, and
  Gleeson}]{MelnikPREunreasonable}
Melnik S., Hackett A., Porter M.~A., Mucha P.~J., Gleeson J.~P. (2011) The unreasonable effectiveness of tree-based theory for networks with clustering. Physical Review E 83(3),~036112.

\bibitem[{Radicchi and Castellano(2016)}]{RadicchiPREbeyond}
Radicchi F., Castellano C. (2016) Beyond the locally treelike approximation for percolation on real networks. Physical Review E 93(3),~030302.

\bibitem[{Ikeda et~al(2010)Ikeda, Hasegawa, and Nemoto}]{Ikeda2010JP}
Ikeda Y., Hasegawa T., Nemoto K. (2010) Cascade dynamics on clustered network. Journal of Physics: Conference Series 221(1),~012005.

\bibitem[{Soramaki et~al(2007)Soramäki, Bech, Arnold, Glass, and
  Beyeler}]{Soramaki2007physicaA}
Soram\"{a}ki K., Bech M.~L., Arnold J., Glass R.~J., Beyeler W.~E. (2007) The topology of interbank payment flows. Physica A 379(1),~317--333.

\bibitem[{Bech and Atalay(2010)}]{BechPhysicaA2010}
Bech M.~L., Atalay E. (2010) The topology of the federal funds market. Physica A 389(22),~5223--5246.

\bibitem[{Dodds and Payne(2009)}]{Dodds2009PRE}
Dodds P.~S., Payne J.~L. (2009) Analysis of a threshold model of social contagion on degree-correlated networks. Physical Review E 79(6),~066115.

\bibitem[{Payne et~al(2009)Payne, Dodds, and Eppstein}]{Payne2009PRE}
Payne J., Dodds P., Eppstein M. (2009) {Information cascades on
  degree-correlated random networks}. Physical Review E 80(2), 026125.

\bibitem[{Hurd et~al(2017)Hurd, Gleeson, and Melnik}]{Hurd2017Plosone}
Hurd T.~R., Gleeson J.~P., Melnik S. (2017) A framework for analyzing contagion in assortative banking networks. PLOS ONE 12(2),~1--20.

\bibitem[{Kivel\"{a} et~al(2014)Kivel\"{a}, Arenas, Barthelemy, Gleeson,
  Moreno, and Porter}]{Kivela2014_multilayer_review}
Kivel\"{a} M., Arenas A., Barthelemy M., Gleeson J.~P., Moreno Y., Porter M.~A. (2014) Multilayer networks. Journal of Complex Networks 2(3),~203--271.

\bibitem[{Brummitt and Kobayashi(2015)}]{Brummitt2015PRE}
Brummitt C.~D., Kobayashi T. (2015) Cascades in multiplex financial networks with debts of different seniority. Physical Review E 91(6),~062813.

\bibitem[{Bargigli et~al(2015)Bargigli, Di~Iasio, Infante, Lillo, and
  Pierobon}]{bargigli2015multiplex}
Bargigli L., Di~Iasio G., Infante L., Lillo F., Pierobon F. (2015) The
  multiplex structure of interbank networks. Quantitative Finance
  15(4),~673--691.

\bibitem[{Poledna et~al(2015)Poledna, Molina-Borboa, Martínez-Jaramillo,
  van~der Leij, and Thurner}]{Polenda2015JFS}
Poledna S., Molina-Borboa J.~L., Martínez-Jaramillo S., van~der Leij M., Thurner S. (2015) The multi-layer network nature of systemic risk and its implications for the costs of financial crises. Journal of Financial Stability 20,~70--81.

\bibitem[{Beale et~al(2011)Beale, Rand, Battey, Croxson, May, and
  Nowak}]{Beale2011}
Beale N., Rand D.~G., Battey H., Croxson K., May R.~M., Nowak M.~A. (2011)
  Individual versus systemic risk and the regulator's dilemma. Proceedings of the  National
  Academy of Sciences USA 108(31),~12647--12652.

\bibitem[{Huang et~al(2013)Huang, Vodenska, Havlin, and
  Stanley}]{huang2013cascading}
Huang X., Vodenska I., Havlin S., Stanley H.~E. (2013) Cascading failures in bi-partite graphs: Model for systemic risk propagation. Scientific Reports 3,~1219.

\bibitem[{Caccioli et~al(2014)Caccioli, Shrestha, Moore, and
  Farmer}]{caccioli2014stability}
Caccioli F., Shrestha M., Moore C., Farmer J.~D. (2014) Stability analysis of financial contagion due to overlapping portfolios. Journal of Banking \& Finance 46,~233--245.

\bibitem[{Caccioli et~al(2015)Caccioli, Farmer, Foti, and
  Rockmore}]{caccioli2015overlapping}
Caccioli F., Farmer J.~D., Foti N., Rockmore D. (2015) Overlapping portfolios, contagion, and financial stability. Journal of Economic Dynamics and Control 51,~50--63.

\bibitem[{Kobayashi(2013)}]{Kobayashi2013EPJB}
Kobayashi T. (2013) Network versus portfolio structure in financial systems. European Physical Journal B 86(10),~434.

\bibitem[{Kobayashi(2014)}]{Kobayashi2014EL}
Kobayashi T. (2014) {A model of financial contagion with variable asset returns may be replaced with a simple threshold model of cascades}. Economics Letters 124,~113--116.

\bibitem[{Glasserman and Young(2015)}]{glasserman2015likely}
Glasserman P., Young H.~P. (2015) How likely is contagion in financial
  networks? Journal of Banking \& Finance 50,~383--399.

\bibitem[{Battiston et~al(2016)Battiston, D’Errico, and
  Visentin}]{battiston2016rethinking}
Battiston S., D'Errico M., Visentin G. (2016) Rethinking financial contagion. arXiv:1608.07831.

\bibitem[{Upper(2011)}]{upper2011simulation}
Upper C. (2011) Simulation methods to assess the danger of contagion in interbank markets. Journal of Financial Stability 7(3),~111--125.

\bibitem[{Glasserman and Young(2015)}]{glasserman2015financial}
Glasserman P., Young H.~P. (2015) Financial networks. Department of Economics Discussion Paper 753, University of Oxford.

\bibitem[{Battiston et~al(2016)Battiston, Caldarelli, D’Errico, and
  Gurciullo}]{battiston2016leveraging}
Battiston S., Caldarelli G., D'Errico M., Gurciullo S. (2016) Leveraging the network: A stress-test framework based on DebtRank. Statistics \& Risk Modeling 33(3-4),~117--138.

\bibitem[{Bardoscia et~al(2015)Bardoscia, Battiston, Caccioli, and
  Caldarelli}]{bardoscia2015debtrank}
Bardoscia M., Battiston S., Caccioli F., Caldarelli G. (2015) DebtRank: A microscopic foundation for shock propagation. PLOS ONE 10(6),~e0130406.

\bibitem[{Bardoscia et~al(2016)Bardoscia, Caccioli, Perotti, Vivaldo, and Caldarelli}]{bardoscia2016distress}
Bardoscia M., Caccioli F., Perotti J.~I., Vivaldo G., Caldarelli G. (2016)
  Distress propagation in complex networks: the case of non-linear DebtRank.
  PLOS ONE 11(10),~e0163825.

\bibitem[{Bardoscia et~al(2017)Bardoscia, Battiston, Caccioli, and
  Caldarelli}]{bardoscia2017pathways}
Bardoscia M., Battiston S., Caccioli F., Caldarelli G. (2017) Pathways towards
  instability in financial networks. Nature Communications 8,~14416.

\bibitem[{Thurner and Poledna(2013)}]{thurner2013debtrank}
Thurner S., Poledna S. (2013) Debtrank-transparency: Controlling systemic risk
  in financial networks. Scientific Reports 3, 1888.

\bibitem[{Poledna and Thurner(2016)}]{poledna2016elimination}
Poledna S., Thurner S. (2016) Elimination of systemic risk in financial networks by means of a systemic risk transaction tax. Quantitative Finance 16(10),~1599--1613.

\bibitem[{Cifuentes et~al(2005)Cifuentes, Ferrucci, and
  Shin}]{cifuentes2005liquidity}
Cifuentes R., Ferrucci G., Shin H.~S. (2005) Liquidity risk and contagion. Journal of the European Economic Association 3(2-3),~556--566.

\bibitem[{Nier et~al(2007)Nier, Yang, Yorulmazer, and
  Alentorn}]{nier2007network}
Nier E., Yang J., Yorulmazer T., Alentorn A. (2007) Network models and
  financial stability. Journal of Economic Dynamics and Control
  31(6),~2033--2060.

\bibitem[{May and Arinaminpathy(2010)}]{may2010systemic}
May R.~M., Arinaminpathy N. (2010) Systemic risk: The dynamics of model banking systems. Journal of the Royal Society Interface 7(46),~823--838.

\bibitem[{Banwo et~al(2016)Banwo, Caccioli, Harrald, and
  Medda}]{banwo2016effect}
Banwo O., Caccioli F., Harrald P., Medda F. (2016) The effect of heterogeneity on financial contagion due to overlapping portfolios. Advances in Complex Systems 19(08),~1650016.

\bibitem[{Greenwood et~al(2015)Greenwood, Landier, and
  Thesmar}]{greenwood2015vulnerable}
Greenwood R., Landier A., Thesmar D. (2015) Vulnerable banks. Journal of Financial Economics 115(3),~471--485.

\bibitem[{Duarte and Eisenbach(2015)}]{duarte2015fire}
Duarte F., Eisenbach T.~M. (2015) Fire-sale spillovers and systemic risk. Staff Report no.645, Federal Reserve Bank of New York.

\bibitem[{Cont and Schaanning(2017)}]{cont2017fire}
Cont R., Schaanning E.~F. (2017) Fire sales, indirect contagion and systemic stress testing. Working Paper 2/2017, Norges Bank. 

\bibitem[{Corsi et~al(2016)Corsi, Marmi, and Lillo}]{corsi2016when}
Corsi F., Marmi S., Lillo F. (2016) When micro prudence increases macro risk:
  The destabilizing effects of financial innovation, leverage, and
  diversification. Operations Research 64(5),~1073--1088.

\bibitem[{Shin(2010)}]{shin2010risk}
Shin H.~S. (2010) \emph{Risk and Liquidity}, Oxford University Press, Oxford.

\bibitem[{Upper and Worms(2004)}]{Upper2004EER}
Upper C., Worms A. (2004) Estimating bilateral exposures in the German
  interbank market: Is there a danger of contagion? European Economic Review 48(4),~827--849.

\bibitem[{Degryse et~al(2007)Degryse, Nguyen et~al}]{Degryse2007IJCB}
Degryse H., Nguyen G., et~al (2007) Interbank exposures: An empirical
  examination of contagion risk in the Belgian banking system. International Journal of Central Banking 3(2),~123--171.

\bibitem[{Lelyveld and Liedorp(2006)}]{Lelyveld2006IJCB}
van Lelyveld I., Liedorp F. (2006) Interbank contagion in the Dutch banking
  sector: A sensitivity analysis. International Journal of Central Banking 2(2),~99--133.

\bibitem[{Wells(2004)}]{wells2004financial}
Wells S. (2004) Financial interlinkages in the United Kingdom's interbank
  market and the risk of contagion. Bank of England Quarterly Bulletin
  44(3),~331.

\bibitem[{Langfield et~al(2014)Langfield, Liu, and Ota}]{langfield2014mapping}
Langfield S., Liu Z., Ota T. (2014) Mapping the UK interbank system. Journal of Banking \& Finance 45,~288--303.

\bibitem[{Furfine(2003)}]{furfine2003interbank}
Furfine C. (2003) Interbank exposures: Quantifying the risk of contagion. Journal of Money, Credit, and Banking 35(1),~111--128.

\bibitem[{Martinez-Jaramillo et~al(2014)Martinez-Jaramillo,
  Alexandrova-Kabadjova, Bravo-Benitez, and
  Sol{\'o}rzano-Margain}]{martinez2014empirical}
Martinez-Jaramillo S., Alexandrova-Kabadjova B., Bravo-Benitez B.,
  Sol{\'o}rzano-Margain J.~P. (2014) An empirical study of the Mexican banking
  system’s network and its implications for systemic risk. Journal of
  Economic Dynamics and Control 40,~242--265.

\bibitem[{Imakubo et~al(2010)Imakubo, Soejima et~al}]{Imakubo2010BOJ}
Imakubo K., Soejima Y., et~al (2010) The transaction network in Japan's interbank money markets. Monetary and Economic Studies 28,~107--150.

\bibitem[{Mistrulli(2011)}]{Mistrulli2011JBF}
Mistrulli P.~E. (2011) Assessing financial contagion in the interbank market:
  Maximum entropy versus observed interbank lending patterns. Journal of Banking \& Finance 35(5),~1114--1127.

\bibitem[{Mastrandrea et~al(2014)Mastrandrea, Squartini, Fagiolo, and
  Garlaschelli}]{mastrandrea2014enhanced}
Mastrandrea R., Squartini T., Fagiolo G., Garlaschelli D. (2014) Enhanced
  reconstruction of weighted networks from strengths and degrees. New Journal of Physics 16(4),~043022.

\bibitem[{Anand et~al(2017)Anand, van Lelyveld, Ádám Banai, Friedrich,
  Garratt, Hałaj, Fique, Hansen, Jaramillo, Lee, Molina-Borboa, Nobili, Rajan,
  Salakhova, Silva, Silvestri, and de~Souza}]{ANAND2017}
Anand K., van Lelyveld I., Ádám Banai, Friedrich S., Garratt R., Hałaj G.,
  Fique J., Hansen I., Jaramillo S.~M., Lee H., Molina-Borboa J.~L., Nobili S.,
  Rajan S., Salakhova D., Silva T.~C., Silvestri L., de~Souza S. R.~S. (2017)
  The missing links: A global study on uncovering financial network structures
  from partial data. Journal of Financial Stability, in press.

\bibitem[{Demiralp et~al(2006)Demiralp, Preslopsky, and
  Whitesell}]{demiralp2006overnight}
Demiralp S., Preslopsky B., Whitesell W. (2006) Overnight interbank loan
  markets. Journal of Economics and Business 58(1),~67--83.

\bibitem[{Borgatti and Everett(2000)}]{borgatti2000coreperiphery}
Borgatti S.~P., Everett M.~G. (2000) Models of core/periphery structures.
  Social Networks 21(4),~375--395.

\bibitem[{Craig and Von~Peter(2014)}]{craig2014interbank}
Craig B., von~Peter G. (2014) Interbank tiering and money center banks. Journal
  of Financial Intermediation 23(3),~322--347.

\bibitem[{Fricke and Lux(2015)}]{Fricke2015CompEcon}
Fricke D., Lux T. (2015) Core--periphery structure in the overnight money
  market: evidence from the e-mid trading platform. Computational Economics
  45(3),~359--395.

\bibitem[{Barucca and Lillo(2016)}]{Barucca2016Soliton}
Barucca P., Lillo F. (2016) Disentangling bipartite and core-periphery
  structure in financial networks. Chaos, Solitons \& Fractals 88,~244--253.

\bibitem[{Barucca and Lillo(2017)}]{Barucca2017Comp}
Barucca P., Lillo F. (2017) The organization of the interbank network and how
  ECB unconventional measures affected the e-MID overnight market.
  Computational Management Science, in press.

\bibitem[{Peixoto(2017)}]{peixoto2017bayesian}
Peixoto T.~P. (2017) Bayesian stochastic blockmodeling. arXiv:1705.10225.

\bibitem[{Rombach et~al(2014)Rombach, Porter, Fowler, and
  Mucha}]{rombach2014core}
Rombach M.~P., Porter M.~A., Fowler J.~H., Mucha P.~J. (2014) Core-periphery
  structure in networks. SIAM Journal on Applied mathematics 74(1),~167--190.

\bibitem[{Kojaku and Masuda(2017{\natexlab{a}})}]{kojaku2017finding}
Kojaku S., Masuda N. (2017{\natexlab{a}}) Finding multiple core-periphery pairs
  in networks. arXiv:1702.06903.

\bibitem[{Kojaku and Masuda(2017{\natexlab{b}})}]{kojaku2017core}
Kojaku S., Masuda N. (2017{\natexlab{b}}) Core-periphery structure requires
  something else in the network. arXiv:1710.07076.

\bibitem[{Kamada and Kawai(1989)}]{Kamada1989algorithm}
Kamada T., Kawai S. (1989) An algorithm for drawing general undirected graphs.
  Information Processing Letters 31(1),~7--15.

\bibitem[{van Lelyveld et~al(2014)}]{van2014finding}
Veld D. in 't, van Lelyveld I. (2014) Finding the core: Network structure in interbank
  markets. Journal of Banking \& Finance 49,~27--40.

\bibitem[{Finger et~al(2013)Finger, Fricke, and Lux}]{finger2013network}
Finger K., Fricke D., Lux T. (2013) Network analysis of the e-MID overnight
  money market: the informational value of different aggregation levels for
  intrinsic dynamic processes. Computational Management Science
  10(2-3),~187--211.

\bibitem[{Musmeci et~al(2013)Musmeci, Battiston, Caldarelli, Puliga, and
  Gabrielli}]{Musmeci2013JSP}
Musmeci N., Battiston S., Caldarelli G., Puliga M., Gabrielli A. (2013)
  Bootstrapping topological properties and systemic risk of complex networks
  using the fitness model. Journal of Statistical Physics 151(3-4),~720--734.

\bibitem[{Estrada and Rodr\'{\i}guez-Vel\'azquez(2005)}]{Estrada2005PRE}
Estrada E., Rodr\'{\i}guez-Vel\'azquez J.~A. (2005) Spectral measures of
  bipartivity in complex networks. Physical Review E 72(4),~046105.

\bibitem[{Cattuto et~al(2010)Cattuto, Van~den Broeck, Barrat, Colizza, Pinton,
  and Vespignani}]{Cattuto2010PLOS}
Cattuto C., Van~den Broeck W., Barrat A., Colizza V., Pinton J.-F., Vespignani
  A. (2010) Dynamics of person-to-person interactions from distributed RFID
  sensor networks. PLOS ONE 5(7),~1--9.

\bibitem[{Starnini et~al(2013)Starnini, Baronchelli, and
  Pastor-Satorras}]{Starnini2013PRL}
Starnini M., Baronchelli A., Pastor-Satorras R. (2013) Modeling human dynamics
  of face-to-face interaction networks. Physical Review Letters 110(16),~168701.

\bibitem[{Schl{\"a}pfer et~al(2014)Schl{\"a}pfer, Bettencourt, Grauwin,
  Raschke, Claxton, Smoreda, West, and Ratti}]{Schlapfer2014}
Schl{\"a}pfer M., Bettencourt L.~M., Grauwin S., Raschke M., Claxton R.,
  Smoreda Z., West G.~B., Ratti C. (2014) The scaling of human interactions
  with city size. Journal of the Royal Society Interface 11(98),~20130789.

\end{thebibliography}

%\begin{thebibliography}{}
%\end{thebibliography}

\end{document}